\documentclass[11pt]{article}

\usepackage{graphicx}
\usepackage{amsmath}
\usepackage{amsfonts}
\usepackage{amssymb}
\usepackage{graphicx}
\usepackage{caption2}

\newcommand{\be}{\begin{equation}}
\newcommand{\ee}{\end{equation}}

\newcommand{\bea}{\begin{eqnarray}}
\newcommand{\eea}{\end{eqnarray}}

\newcommand{\bi}{\begin{itemize}}
\newcommand{\ei}{\end{itemize}}

\newcommand{\ben}{\begin{enumerate}}
\newcommand{\een}{\end{enumerate}}

\newcommand{\bef}{\begin{figure}[tbp]}
\newcommand{\enf}{\end{figure}}

\newcommand{\bt}{\begin{tabular}{lcllcl}}
\newcommand{\et}{\end{tabular}}

\newcommand{\bd}{\begin{description}}
\newcommand{\ed}{\end{description}}

\newtheorem{theorem}{Theorem}
\newtheorem{lemma}{Lemma}[section]
\newtheorem{corollary}{Corollary}

\newcounter{example}

\newenvironment{example}
 {\refstepcounter{example}%
  \vspace{.25cm}%
  \noindent%
  {\bf \boldmath Example \arabic{example}:}%
  \noindent}
 {\hfill $\Box$ }

\newenvironment{proof}[1]
 {\noindent%
 {\bf \boldmath Proof #1:}}
 {\hfill $\Box$  \\}

\newcommand{\eref}[1]{(\ref{#1})}       

\newcommand{\dfn}{\stackrel{\triangle}{=}}  

\newcommand{\comb}[2]{\left (
 \raisebox{-4pt}{$\stackrel{\mbox{\large $#1$}}{#2}$} \right ) }

\newcommand{\kvec} {{\mathbf k}}
\newcommand{\cvec} {{\mathbf c}}

\newcommand{\pvec}   {\mbox{\boldmath $\theta$}}
\newcommand{\psivec} {\mbox{\boldmath $\psi$}}
\newcommand{\sigvec} {\mbox{\boldmath $\sigma$}}

\newcommand{\tauvec} {\mbox{\boldmath $\tau$}}
\newcommand{\etavec} {\mbox{\boldmath $\eta$}}
\newcommand{\xivec}  {\mbox{\boldmath $\xi$}}
\newcommand{\kappavec}  {\mbox{\boldmath $\kappa$}}

\newcommand{\pvece}{\hat{\pvec}}

\newcommand{\Beta}{{\cal B}}

\newcommand{\eventA}{{\cal F}}

\newcommand{\Sset}{{\cal S}}
\newcommand{\Tset}{{\cal T}}

\begin{document}

\title{Patterns of i.i.d.\ Sequences and Their Entropy - Part I: General Bounds\footnote{Supported
in part by NSF Grant CCF-0347969.  Parts of the material in this
paper were presented at the 41st and 42nd Annual Allerton
Conferences on Communication, Control, and Computing, Monticello,
IL, October 2003, September-October 2004, and at the IEEE
Information Theory Workshop on Coding and Complexity, Rotorua, New
Zealand, August-September, 2005.}}
\author{Gil I. Shamir \\
 Department of Electrical and Computer Engineering \\
 University of Utah \\
 Salt Lake City, UT 84112, U.S.A \\
 e-mail: gshamir@ece.utah.edu.}
\date{}
\maketitle

\begin{abstract}
Tight bounds on the block entropy of \emph{patterns\/} of sequences
generated by independent and identically distributed (i.i.d.)
sources are derived.  A pattern of a sequence is a sequence of
integer indices with each index representing the order of first
occurrence of the respective symbol in the original sequence.  Since
a pattern is the result of data processing on the original sequence,
its entropy cannot be larger.  Bounds derived here describe the
pattern entropy as function of the original i.i.d.\ source entropy,
the alphabet size, the symbol probabilities, and their arrangement
in the probability space.  Matching upper and lower bounds derived
provide a useful tool for very accurate approximations of pattern
block entropies for various distributions, and for assessing the
decrease of the pattern entropy from that of the original i.i.d.\
sequence.

{\bf Index Terms}: patterns, index sequences, entropy.
\end{abstract}

\section{Introduction}
\label{sec:introduction}

Several recent works (see, e.g., \cite{aberg97}, \cite{jevtic02},
\cite{orlitsky04}, \cite{shamir03},
\cite{shamir04c}, \cite{shamir04d}) have considered universal
compression for \emph{patterns\/} of independent and identically
distributed (i.i.d.)\ sequences. The pattern of a sequence
$x^n \dfn \left ( x_1, x_2, \ldots, x_n \right )$
is a sequence
$\psi^n \dfn \psivec \dfn \Psi \left ( x^n \right )$
of pointers that point to the actual alphabet letters,
where the alphabet letters are assigned \emph{indices\/} in order of
first occurrence.
For example, the pattern of all sequences $x^n = lossless$,
$x^n = sellsoll$, $x^n = 12331433$, and $x^n = 76887288$
is $\psi^n =\Psi \left ( x^n \right ) = 12331433$.  Capital $\Psi (\cdot)$ is used
to denote the operator of taking a pattern of a sequence.
A pattern sequence thus contains all positive integers
from $1$ up to a maximum value in increasing order of first
occurrence, and is also independent of the alphabet of the actual
data.

Universal compression of patterns is interesting when
compressing sequences generated by an
initially unknown alphabet, such as a document in an unknown
language.  In such applications, separate dictionary and pattern compression can
be performed.  Most initial work on this topic focused on showing diminishing
universal compression redundancy rates for the \emph{individual sequence\/} case
\cite{jevtic02}, \cite{orlitsky04}, and for the average case
\cite{shamir03}, \cite{shamir04c}, \cite{shamir04d}.
However, since a pattern $\Psi \left (x^n \right )$ is the result of data processing on
the original sequence $x^n$, its entropy must be no greater than that of the original sequence.
Specifically, if $x^n$ is generated by an i.i.d.\ source of alphabet size $k$,
\be
 \label{eq:entropy_bounds_simple}
 nH_{\theta} \left (X \right ) - \log \left [ k!/ \left (\max \left \{ 0, k-n \right \} \right)! \right ]  \leq
 H_{\theta} \left ( \Psi^n \right ) \leq
 nH_{\theta} \left (X \right ),
\ee
where capital letters denote random variables, and $\pvec$ is the probability parameter
vector governing the source.  The lower bound is
since $H_{\theta} \left ( \Psi^n \right ) =
H_{\theta} \left (X^n, \Psi^n \right ) - H_{\theta} \left (X^n | \Psi^n \right ) =
H_{\theta} \left (X^n \right ) - H_{\theta} \left (X^n | \Psi^n \right )$, where the
second equality is because there is no uncertainty about $\Psi^n$ given $X^n$.  Finally,
$H_{\theta} \left (X^n | \Psi^n  \right )$ is bounded by logarithm\footnote{Logarithms are
taken to base $2$, here and elsewhere.  The natural logarithm is denoted by $\ln$.} of the total
possible mappings from indices to symbols.

The bounds in \eref{eq:entropy_bounds_simple} already show that
for $k = o(n)$, the pattern entropy rate equals the i.i.d.\ one
for non-diminishing $H_{\theta}(X)$.
However, the bounds in \eref{eq:entropy_bounds_simple} are usually loose.
Specifically, the
\emph{description length\/} shown for sufficiently large alphabets
in \cite{shamir03} (see also \cite{shamir04d})
for a \emph{universal\/} sequential compression method for patterns
was significantly smaller than the block i.i.d.\ entropy.
This indicates that not only is there an entropy decrease
in patterns, but for large alphabets, this decrease is more significant than
universal coding redundancy.
Hence, it is essential to study the behavior of the pattern entropy.
Pattern entropy is also important in
learning applications.   Consider all the new species an explorer observes.  The explorer
can identify these species with the first time each was seen.  There is no difference
if it sees specie $A$ or specie $B$ (and never sees the other).  The next time the observed
specie is seen, it is identified with its index.
The entropy of patterns
can model uncertainty of such processes.
Its exponent gives an approximate count of the typical patterns
one is likely to observe.
If the uncertainty goes to $0$, we are likely to observe only one pattern.

Initial results from this paper, first presented in
\cite{shamir03a}, bounded the range of values within which
the entropy of a pattern can be, depending on the alphabet size.
Subsequently to our initial results \cite{shamir03a},
pattern entropy \emph{rates\/} were independently studied
with a different view of the problem
in \cite{gemelos06} and \cite{orlitsky06}.
The main result was that for discrete i.i.d.\ sources the pattern
entropy rate is
equal to that of the underlying i.i.d.\ process.
This result was also extended to discrete finite entropy stationary processes.
Some limiting \emph{order of magnitude\/} bounds on \emph{block\/}
pattern entropies were also provided.

This paper extensively studies block entropy of patterns, providing
tight upper and matching lower bounds on the block entropy.  The bound
pairs can be used together to provide very accurate approximations
of the entropy of $\Psi^n$.  Specific
distributions are studied in \cite{shamir07}.
The basic method partitions
the probability space into a \emph{grid\/} of points.
Between each two points, we obtain a \emph{bin\/}.
Symbols whose probabilities lie in the same bin can be exchanged in a given
$x^n$ to provide another sequence $x'^n$ with the same pattern and almost equal probability.
Counting all these sequences leads to the bounds on the pattern entropy.
Very low probabilities are combined into one point mass.
A key factor in obtaining tight bounds is a proper choice of increased-spacing grids.

The outline of the paper is as follows.
Section~\ref{sec:note_def} defines some notation and presents some preliminaries.
A summary of the main results in the paper is given in Section~\ref{sec:main_results}.
Then, in Section~\ref{sec:large}, upper and lower bounds
for pattern entropy of i.i.d.\ sources with sufficiently large probabilities are derived.
Section~\ref{sec:very} contains the derivations of more general
upper and lower bounds, that do not require a condition on the letter probabilities.
Finally, Section~\ref{sec:large_range}
shows the range of values that the pattern entropy can take for bounded probabilities, depending
on the actual source distribution.

\section{Preliminaries}
\label{sec:note_def}

Let $x^n$ be an $n$-tuple with components
$x_i \in \Sigma \dfn \left \{1, 2, \ldots, k \right \}$ (where the alphabet is defined without
loss of generality).
The asymptotic regime is that $n \rightarrow \infty$, but $k$ may
also be greater then $n$.
The vector $\pvec \dfn \left ( \theta_1, \theta_2, \ldots, \theta_k \right )$
is the set of probabilities of all letters in $\Sigma$.
Since the order of the probabilities does not affect the
pattern, we assume, without loss
of generality, that
$\theta_1 \leq \theta_2 \leq \cdots \leq \theta_k$.
Boldface letters denote vectors, whose components
are denoted by their indices.
Capital letters will denote random variables.
The probability of $\psi^n$ \emph{induced\/} by an i.i.d.\ source is
\be
\label{eq:pattern_probability}
 P_{\theta} \left ( \psi^n \right ) =
  \sum_{y^n: \Psi (y^n) = \psi^n } P_{\theta} \left ( y^n \right ).
\ee
This probability can also be expressed
by fixing the actual sequence and summing over all permutations of
occurring symbols of the parameter vector, i.e.,
\be
 \label{eq:pattern_prob1}
 P_{\theta} \left [ \Psi \left ( x^n \right ) \right ] =
 \sum_{\sigvec = \left \{ \sigma_i, i \in x^n \right \}} P_{\theta(\sigma)} \left (x^n \right ),
\ee
where $\sigvec = \left \{ \sigma_1, \ldots, \sigma_k \right \}$ is a permutation set.
For example, if $\pvec = \left ( 0.4, 0.1, 0.2, 0.3 \right )$ and
$\sigvec = \left (3, 1, 4, 2 \right )$,
then $\pvec \left ( \sigvec \right ) = \left ( 0.2, 0.4, 0.3, 0.1 \right )$
and $\theta \left ( \sigma_2 \right ) = \theta_1 = 0.4$.
The only relevant components of $\sigvec$ in \eref{eq:pattern_prob1} are those
of occurring symbols.  Thus if only $m < k$ symbols occur in $x^n$, there
are only $k!/(k - m)!$ elements in the sum in \eref{eq:pattern_prob1}.
The entropy rate of an i.i.d.\ source is
$H_{\theta} \left (X \right )$, and its sequence (block) entropy
is $H_{\theta} \left (X^n \right ) = n H_{\theta} \left (X \right )$.
The \emph{pattern sequence entropy\/} of order $n$ is
\be
\label{eq:pattern_entropy}
 H_{\theta} \left ( \Psi^n \right ) \dfn
 -\sum_{\psi^n} P_{\theta} \left ( \psi^n \right )
 \log P_{\theta} \left ( \psi^n \right ).
\ee

To derive bounds on the pattern entropy, we define three different grids: $\tauvec$,
$\etavec$, and $\xivec$, the first two for upper bounding and the third for lower bounding.
Spacing between grid points is motivated by the fact that
two probability parameters $\theta$ and $\theta'$
separated by $O\left (\sqrt{\theta} / \sqrt{n}^{1+\delta}\right )$; $\delta > 0$,
are near enough to
appear similar in $x^n$.  On the other hand, if
$\left | \theta - \theta' \right | > \sqrt{\theta}/\sqrt{n}^{1-\delta}$,
the parameters are far enough to appear different.
For simplicity of notation, we omit the dependence on $n$ from definitions of grid points.
For $\varepsilon > 0$, let $\tauvec \dfn \left (\tau_0, \tau_1,
\tau_2, \ldots, \tau_b, \ldots, \tau_{B_{\tau}} \right )$ be a grid of $B_{\tau}+1$
points defined by
$\tau_0 = 0$, and
\be
 \label{eq:tau_grid_def}
 \tau_b =
 \sum_{j=1}^{b} \frac{2 (j - \frac{1}{2})}{n^{1+\varepsilon}} =
 \frac{b^2}{n^{1+\varepsilon}},~~
  b = 1, 2, \ldots, B_{\tau}.
\ee
Let $\etavec'$ be defined almost like $\tauvec$,
\be
 \label{eq:eta_grid_def2}
 \eta'_b \dfn
 \sum_{j=1}^{b} \frac{2 (j - \frac{1}{2})}{n^{1+2\varepsilon}} =
 \frac{b^2}{n^{1+2\varepsilon}}.
\ee
The grid $\etavec \dfn \left ( \eta_0, \eta_1, \ldots, \eta_{B_{\eta}} \right )$ is
defined by $\eta_0 = 0$, $\eta_1 = \tau_1 = \frac{1}{n^{1+\varepsilon}}$,
$\eta_2 = \frac{1}{n^{1-\varepsilon}}$, and
\be
\label{eq:eta_grid_def}
 \eta_b = \eta'_{b+\left \lfloor n^{3\varepsilon/2} \right \rfloor - 2},~~
  b = 3,4, \ldots, B_{\eta}.
\ee
Unlike $\tauvec$ and $\etavec$,
$\xivec \dfn \left ( \xi_0, \xi_1, \ldots, \xi_{B_{\xi}} \right )$ is
defined for lower bounds purposes.  It is defined in a similar manner as the others, where
$\xi_0 = 0$, and for an arbitrarily small $\varepsilon > 0$,
\be
 \label{eq:xi_grid_def}
 \xi_b \dfn \sum_{j=1}^b \frac{2(j-0.5)}{n^{1-\varepsilon}} =
 \frac{b^2}{n^{1-\varepsilon}},~~ b = 1, 2, \ldots, B_{\xi}.
\ee
For all grids, $\tau_{B_{\tau}+1} = \eta_{B_{\eta}+1} = \xi_{B_{\xi}+1} \dfn 1$.
We thus have
$B_{\tau} = \left \lfloor \sqrt{n}^{1+\varepsilon} \right \rfloor$,
$B_{\eta} = \left \lfloor \sqrt{n}^{1+2\varepsilon} \right \rfloor -
\left \lfloor n^{3\varepsilon/2} \right \rfloor + 2$, and
$B_{\xi} = \left \lfloor \sqrt{n}^{1-\varepsilon} \right \rfloor$.
We also define the maximal indices $A_{\tau}$, $A_{\eta}$, and $A_{\xi}$ whose
grid points do not exceed $0.5$ for $\tauvec$, $\etavec$, and $\xivec$, respectively.
Hence,
$A_{\tau} = \left \lfloor \sqrt{n}^{1+\varepsilon}/\sqrt{2} \right \rfloor$,
$A_{\eta} = \left \lfloor \sqrt{n}^{1+2\varepsilon}/ \sqrt{2} \right \rfloor -
\left \lfloor n^{3\varepsilon/2} \right \rfloor + 2$, and
$A_{\xi} = \left \lfloor \sqrt{n}^{1-\varepsilon} /\sqrt{2} \right \rfloor$.

By definition of $\etavec$, for every
$\theta \in\left [\eta_b, \eta_{b+1} \right ]$ where $b \geq 2$,
\be
 \label{eq:tau_grid_spacing}
 \eta_{b+1} - \eta_b =
 \frac{2\left [ (b+d) + 0.5\right ]}{n^{1+2\varepsilon}} \leq
 \frac{3(b+d)}{n^{1+2\varepsilon}}
 = \frac{3 \sqrt{\eta'_{b+d}}}{\sqrt{n}^{1+2\varepsilon}}
 = \frac{3 \sqrt{\eta_b}}{\sqrt{n}^{1+2\varepsilon}}
 \leq \frac{3\sqrt{\theta}}{\sqrt{n}^{1+2\varepsilon}},
\ee
where $d \dfn \left \lfloor n^{\varepsilon/2} \right \rfloor - 1$.
A similar bound applies to $\tau_b$, $b \geq 1$, with $\varepsilon$ in place of $2\varepsilon$.
Similarly,
\be
 \label{eq:xi_grid_spacing}
 \xi_{b+1} - \xi_b =
 \frac{2(b+0.5)}{n^{1-\varepsilon}} =
 \frac{2 \left ( \sqrt{\xi_b} \sqrt{n}^{1 - \varepsilon} + 0.5 \right )}{n^{1-\varepsilon}}
 \geq
 \frac{2\sqrt{\xi_b}}{\sqrt{n}^{1-\varepsilon}}.
\ee

We use $c_b$; $b=0, 1, \ldots, B_{\tau}$, $k_b$; $b=0, 1, \ldots, B_{\eta}$, and $\kappa_b$,
$b=0, 1, \ldots, B_{\xi}$, to denote the number of symbols
for which
$\theta_i \in \left (\tau_b, \tau_{b+1} \right ]$,
$\theta_i \in \left (\eta_b, \eta_{b+1} \right ]$,
and
$\theta_i \in \left (\xi_b, \xi_{b+1} \right ]$, respectively.
Respective vectors containing all components are denoted by
$\cvec$, $\kvec$, and $\kappavec$.  In addition, define
$\kappa'_b$; $b = 1, 2,\ldots, B_{\xi}$, as zero
if $\kappa_b$ is zero, and otherwise, as the number of symbols for which
$\theta_i \in \left (\xi_{b-1}, \xi_{b+2} \right ]$,
with the exception of $\kappa'_1$,
which will only count letters for which
$\theta_i \in \left (\xi_1, \xi_3 \right ]$.
(There is clearly
an overlap between adjacent counters in $\kappavec'$, which is needed for derivation
of a lower bound.)

The grid $\tauvec$ is defined so that all letters
$\theta_i \leq 1/n^{1+\varepsilon}$
are grouped in the same bin.
Grid $\etavec$ also groups probabilities in
$\left ( 1/n^{1+\varepsilon}, 1/n^{1-\varepsilon} \right ]$ in bin $1$.
In particular, $k_0$ and $k_1$ denote the symbol counts of the two groups, respectively.
We will also use $k_{01} \dfn k_0 + k_1$ to denote
the total letters with
$\theta_i \leq 1/n^{1-\varepsilon}$ (thus $k-k_{01}$ denotes the count of symbols with
$\theta_i > 1/n^{1-\varepsilon}$).
Let
\be
 \varphi_b \dfn \sum_{\theta_i \in \left (\eta_b, \eta_{b+1} \right ]} \theta_i
\ee
be the total probability of letters in bin $b$ of grid
$\etavec$.
Of particular importance will be $\varphi_0$,
$\varphi_1$, defined with respect to (w.r.t.)\ bins $0$, $1$, respectively, and
$\varphi_{01} \dfn \varphi_0 + \varphi_1$.  Define
$\ell_0$, $\ell_1$, and $\ell_{01}$, where $\ell_b \dfn \min
\left (k_b, n \right )$.

The probability that letter $i$ does not occur in $X^n$ is
\be
 \label{eq:no_letter_i}
 P_{\theta} \left ( i \not \in X^n \right ) =
 \left ( 1 - \theta_i \right )^n.
\ee
Taking an exponent of the logarithm of \eref{eq:no_letter_i}, using
Taylor series expansion in the exponent,
\be
 \label{eq:no_letter_i_bound}
  e^{-n \left (\theta_i + \theta_i^2 \right )} \leq
  P_{\theta} \left ( i \not \in X^n \right ) \leq
   e^{-n\theta_i}, ~~\mbox{if}~\theta_i \leq 3/5.
\ee
If $\theta_i > 3/5$, the upper bound is the same, but the lower bound is $0$.
Following \eref{eq:no_letter_i_bound},
\be
 \label{eq:yes_letter_i_bound}
 1 -  e^{-n\theta_i} \leq
 P_{\theta} \left ( i \in X^n \right ) \leq
 1 - e^{-n \left (\theta_i + \theta_i^2 \right )},
\ee
where the upper bound is replaced by $1$ for $\theta_i > 3/5$.

The mean number of occurrences of letter $i$ in $X^n$ is given by
$E_{\theta} N_x \left ( i \right ) = n \theta_i$,
where $n_x \left ( i \right )$ is the number of occurrences of $i$ in $x^n$,
$N_x \left ( i \right )$ is its random variable, and $E_{\theta}$ is expectation
given $\pvec$.  Then,
the mean number of re-occurrences (beyond the first occurrence)
of letter $i$ in $X^n$ is given by
\be
 \label{eq:mean_reoccur_i}
 E_{\theta} N_x \left ( i \right ) - P_{\theta} \left ( i \in X^n \right ) =
 n \theta_i - 1 + \left ( 1 - \theta_i \right )^n.
\ee
It is thus bounded by
\be
 \label{eq:mean_reoccur_bound}
 n \theta_i - 1 + e^{-n \left (\theta_i + \theta_i^2 \right )} \leq
 E_{\theta} N_x \left ( i \right ) - P_{\theta} \left ( i \in X^n \right )  \leq
 n \theta_i - 1+ e^{-n\theta_i},
\ee
where, again, the last term of the lower bound is replaced by $0$ for $\theta_i > 3/5$.
Using the Binomial expansion on \eref{eq:no_letter_i},
the probability of an occurrence of
letter $i$ for $\theta_i \leq 1/n$ can be bounded by
\be
 \label{eq:yes_letter_i0}
 n \theta_i - \comb{n}{2} \theta_i^2 \leq
 P_{\theta} \left ( i \in X^n \right ) \leq
 n \theta_i - \comb{n}{2} \theta_i^2 + \comb{n}{3} \theta_i^3,
\ee
and then the mean number of re-occurrences of letter $i$ is bounded by
\be
 \label{eq:mean_recur_bound0}
 \comb{n}{2} \theta_i^2 - \comb{n}{3} \theta_i^3 \leq
 E_{\theta} N_x \left ( i \right ) - P_{\theta} \left ( i \in X^n \right )  \leq
 \comb{n}{2} \theta_i^2.
\ee

Let $K_b$; $C_b$ denote random variables counting the \emph{distinct\/}
symbols from bin $b$ of $\etavec$; $\tauvec$, respectively,
that occur in $X^n$.
Let $K$ denote the total number
of distinct letters occurring in $X^n$.
The mean number of distinct letters from bin $b$ of $\etavec$ that occur in $X^n$ is
\be
 \label{eq:mean_bin}
 L_b \dfn E_{\theta} \left [ K_b \right ]= \sum_{\theta_i \in \left ( \eta_b, \eta_{b+1} \right ]}
 \left [ 1 - \left ( 1 - \theta_i \right )^n \right ]
\ee
where $L_0$, $L_1$, $L_{01}$ are of specific interest, and also $L \dfn E_{\theta} \left [ K \right ]$ is computed
in a similar manner.
As in \eref{eq:yes_letter_i_bound},
\be
 \label{eq:mean_bin_bound}
 k_b - \sum_{\theta_i \in \left ( \eta_b, \eta_{b+1} \right ]}
 e^{-n \theta_i} \leq
 L_b \leq
 k_b - \sum_{\theta_i \in \left ( \eta_b, \eta_{b+1} \right ], ~\theta_i \leq 3/5}
  e^{-n \left (\theta_i+\theta_i^2 \right )}.
\ee
In particular, for bin $0$, as in \eref{eq:yes_letter_i0},
\be
 \label{eq:min_bin0_bound}
 n \varphi_0 - \comb{n}{2} \sum_{i=1}^{k_0}
 \theta_i^2 \leq L_0 \leq
 n \varphi_0 - \comb{n}{2} \sum_{i=1}^{k_0}\theta_i^2 +
 \comb{n}{3} \sum_{i=1}^{k_0}\theta_i^3.
\ee

Packing lower bin(s) into single point masses, we can thus define,
\bea
 \label{eq:zero_bin_packed_entropy}
 H_{\theta}^{(0)} \left ( X \right ) &\dfn&
 -\varphi_0 \log \varphi_0 - \sum_{i=k_0+1}^k \theta_i \log \theta_i, \\
 \label{eq:zeroone_bin_packed_entropy}
 H_{\theta}^{(01)} \left ( X \right ) &\dfn&
 -\varphi_{01} \log \varphi_{01} - \sum_{i=k_{01}+1}^k \theta_i \log \theta_i, \\
 \label{eq:zero_one_bin_packed_entropy}
 H_{\theta}^{(0,1)} \left ( X \right ) &\dfn&
 -\sum_{b=0}^1 \varphi_b \log \varphi_b -
 \sum_{i=k_{01}+1}^k \theta_i \log \theta_i.
\eea
The expressions in \eref{eq:zero_bin_packed_entropy}-\eref{eq:zero_one_bin_packed_entropy}
will be used to express some of the bounds in the paper, where low probability letters
are packed into one or two point masses.  (These expressions also depend on the choice of
$\varepsilon$.  This dependence is omitted for convenience.)

\section{The Main Results}
\label{sec:main_results}

The main results in the paper are summarized below.  First,
if $\theta_i > 1/n^{1-\varepsilon}$, $\forall i$,
the pattern entropy is bounded by
\be
 \label{eq:summary_ublb}
 n H_{\theta} \left ( X \right ) -
 \sum_{b=1}^{A_{\xi}} \log \left ( \kappa_b ! \right ) - k \log 3 - o(1) \leq
 ~H_{\theta} \left ( \Psi^n \right )~ \leq
 n H_{\theta} \left ( X \right ) -
 \left ( 1 - \varepsilon \right )
 \sum_{b=2}^{A_{\eta}} \log \left ( k_b ! \right ) + o(k).
\ee
Namely, the pattern entropy decreases to first order from the i.i.d.\ block entropy
by the logarithm of the product of permutations within
all the bins of the probability space.
The bounds in \eref{eq:summary_ublb} depend on the arrangement of the letters
in the probability space.  However, even if we only know the number of
letters in the alphabet, we can still bound the range that the pattern entropy
can be in.
The actual point in this range does depend on the arrangement of
the letters in the probability space.  However, if the
alphabet is large enough, the pattern entropy \emph{must\/} decrease w.r.t.\
the i.i.d.\ one regardless of this arrangement.  In all, if
$\theta_i > 1/n^{1-\varepsilon}$, $\forall i$, we have
\be
 \label{eq:summary_range}
 n H_{\theta} \left ( X \right ) - \log \left (k! \right ) \leq
 ~H_{\theta} \left ( \Psi^n \right )~
 \leq
 \left \{
 \begin{array}{ll}
  n H_{\theta} \left ( X \right ), &
  \mbox{if}~k < n^{1/3+\varepsilon}, \\
  n H_{\theta} \left ( X \right ) -
  \frac{3}{2} k \log \frac{k}{e n^{1/3+\varepsilon/2}}, &
  \mbox{if}~k \geq n^{1/3+\varepsilon}.
 \end{array}
 \right .
\ee
The bound above shows that the decrease in the pattern entropy w.r.t.\ the
i.i.d.\ one for large alphabets
is to first order between $\log k$ bits and $\log \left ( k^{1.5} / \sqrt{n} \right )$
bits for each alphabet letter.

If the alphabet contains letters with low probabilities, namely, with
$\theta_i \leq 1/n^{1-\varepsilon}$ ($k_{01} > 0$), the pattern entropy is upper bounded by
\bea
 \nonumber
 H_{\theta} \left ( \Psi^n \right ) &\leq&
 n H^{(0,1)}_{\theta} \left ( X \right ) -
 \sum_{b=2}^{A_{\eta}} \left ( 1 - \varepsilon \right )
 \log \left ( k_b ! \right ) \\ &+&
 \nonumber
 \left ( n \varphi_1 - L_1 \right ) \log
 \left [ \min \left \{ k_1, n \right \} \right ] +
 n \varphi_1 h_2 \left ( \frac{L_1}{n\varphi_1} \right ) \\ &+&
 \label{eq:summary_ub3}
 \left ( \frac{n^2}{2} \sum_{i=1}^{k_0} \theta_i^2 \right )
 \log  \left \{ \frac{2 e \cdot \varphi_0 \cdot \min \left \{k_0, n \right \}}
 {n \sum_{i=1}^{k_0} \theta_i^2} \right \},
\eea
where $h_2 \left ( \alpha \right ) \dfn
-\alpha \log \alpha - (1-\alpha) \log (1 - \alpha)$.
The third and forth terms contribute at most
$O \left ( n \varphi_1 \log n \right )$, and the last term $o(n)$.  The bound
in \eref{eq:summary_ub3} implies that the source appears to contain a single letter
for bin $0$ and another single letter for bin $1$, and its entropy decreases, again,
by the logarithm of the number of permutations leading to typical sequences w.r.t.\
all other bins.  In addition, there is a limited penalty reflected in the last three
terms for packing all letters in bins $1$ and $0$ as two point masses.  This penalty is
higher for bin $1$, which is the boundary between two different asymptotic behaviors.
For non-diminishing i.i.d.\ entropies $H^{(0,1)}_{\theta} \left ( X \right )$ the penalty
of packing all letters in bin $0$ into one point mass is negligible.

A lower bound of a similar nature is then obtained, showing that the pattern entropy
satisfies
\bea
 \nonumber
 H_{\theta} \left ( \Psi^n \right ) &\geq&
 n H^{(01)}_{\theta} \left ( X \right ) -
 \sum_{b=1}^{A_{\xi}} \log \left ( \kappa_b ! \right ) - \left (k - \kappa_0 \right ) \log 3 \\
 \nonumber
 &+&
 \sum_{i=1}^{k_{01} - 1} \left [ n \theta_i - 1 + e^{-n \left ( \theta_i +
 \theta_i^2 \right )} \right ]
 \log \frac{\varphi_{01}}{\theta_i} +
 \left ( n \theta_{k_{01}} - 1 \right ) \log \frac{\varphi_{01}}{\theta_{k_{01}}} \\
 &+&
 (\log e) \sum_{i=1}^{L_{01}-1} \left ( L_{01} - i \right )
 \frac{\theta_i}{\varphi_{01}} -
 \log \comb{k_{\vartheta}^- + k_{\vartheta}^+}{k_{\vartheta}^+} - o(1),
 \label{eq:summary_lb2}
\eea
where $k_{\vartheta}^-$ denotes the number of letters
with $\theta_i \in \left ( \vartheta^- / n^{1-\varepsilon}, 1/ n^{1-\varepsilon} \right ]$ and
$k_{\vartheta}^+$ the number of letters with
$\theta_i \in \left ( 1/ n^{1-\varepsilon}, \vartheta^+ / n^{1-\varepsilon} \right ]$,
and $\vartheta^-$ and $\vartheta^+$ are constants, such that $\vartheta^+ > 1 > \vartheta^- > 0$.
This bound illustrates similar behavior to that in \eref{eq:summary_ub3}, where the
pattern entropy behaves like that of a source for which the low probabilities in bins $0$ and $1$
are packed into one point mass, and a similar behavior
to that in \eref{eq:summary_ublb} is shown for greater probabilities.
Packing of bins $01$ results in correction terms
reflecting the increase in entropy due to repetitions and first occurrences, and another correction
term (the seventh term) reflecting the unclear boundary between two different asymptotic behaviors.
For many sources, variations of
the last two bounds are very close to each other and lead to very accurate approximations
of the pattern entropy \cite{shamir07}.

\section{Bounds for Small and Large Alphabets}
\label{sec:large}

This section studies pattern entropy with bounded letter probabilities
$\theta_i > 1/n^{1-\varepsilon},~\forall i$ (i.e., $k_{01} = 0$).
Upper and lower bounds are presented.

\subsection{An Upper Bound}
\label{sec:large_upper}

\begin{theorem}
\label{theorem:ub1}
Fix $\delta > 0$.
Let $n \rightarrow \infty$ and $\varepsilon \geq (1+\delta)(\ln \ln n)/(\ln n)$.
If $\theta_i > 1/n^{1-\varepsilon}$, $\forall i, 1\leq i \leq k$,
\be
 \label{eq:ub1}
 H_{\theta} \left ( \Psi^n \right ) \leq
 n H_{\theta} \left ( X \right ) -
 \left ( 1 - \varepsilon \right )
 \sum_{b=2}^{A_{\eta}} \log \left ( k_b ! \right ) + o(k).
\ee
\end{theorem}
The bound can be tightened by substituting $\varepsilon$ in the second term
by $\exp \left \{-\left [ 0.1n^{\varepsilon} - 2\ln n \right ] \right \}$.
The grid $\etavec$, which is used for the proof, is defined with the same $\varepsilon$.
Theorem~\ref{theorem:ub1} shows
that letters whose probabilities
are in the same bin of $\etavec$
can be exchanged in a typical $x^n$ generating sequences $x'^n$ with
$P_{\theta} \left ( x'^n \right ) \approx P_{\theta} \left ( x^n \right )$ and
$\Psi \left (x'^n \right ) =\Psi \left (x^n \right )$.
This increases $P_{\theta} \left [ \Psi \left (x^n \right ) \right ]$
by a factor of the total of
such possible permutations, and decreases the entropy by its logarithm.
Summation in the second term of \eref{eq:ub1} is only up to $A_{\eta}$
because larger index bins contain at most a single symbol probability.

\begin{proof}{}
The proof separates typical $x^n$ from unlikely (untypical) $x^n$.  Then,
$P_{\theta} \left ( \psi^n \right )$ is lower bounded by the sum
$P_{\theta} \left ( x^n \right )$ of typical $x^n$ with $\Psi \left ( x^n \right ) = \psi^n$.
For all such $x^n$, $P_{\theta} \left (x^n \right )$ is almost equal.
The number of such sequences results in the entropy decrease and is equal to the number of possible
permutations within the bins of $\etavec$.

We define a typical set.
Let $\pvece$ be the \emph{Maximum Likelihood\/} (ML)
estimator of $\pvec$ from $x^n$.  Then,
\bea
 \label{eq:typical_set}
 \Tset_x &\dfn& \left \{ x^n~:~ \forall i,
 \left | \hat{\theta}_i - \theta_i \right | <
 \frac{\sqrt{\theta_i}}{2\sqrt{n}^{1-\varepsilon}} \right \}, \\
 \label{eq:untypical_set}
 \bar{\Tset}_x &=& \left \{ x^n~:~ \exists i,
 \left | \hat{\theta}_i - \theta_i \right | \geq
 \frac{\sqrt{\theta_i}}{2\sqrt{n}^{1-\varepsilon}} \right \}.
\eea
\begin{lemma}
 \label{lemma_untyp_prob}
 \be
  \label{eq:non_typ_prob}
  P_{\theta} \left ( \bar{\Tset}_x \right ) \leq
  \exp \left \{ -0.1 n^{\varepsilon} + \left ( 2 - \varepsilon \right ) \ln n
  \right \} \dfn \varepsilon_n.
 \ee
\end{lemma}
The proof of Lemma~\ref{lemma_untyp_prob}
is in \ref{ap:lemma_untyp_prob_proof}.
For the choice of $\varepsilon$ in 
Theorem~\ref{theorem:ub1},
$\varepsilon_n \rightarrow 0$.

Now, define $\Sset$ as the set of all permutations
$\sigvec$ that permute symbols only within bins of $\etavec$, i.e.,
\be
 \label{eq:Sset_def}
 \Sset \dfn \left \{ \sigvec ~:~
 \theta_i \in \left ( \eta_b, \eta_{b+1} \right ]~\Rightarrow~
 \theta(\sigma_i) \in \left ( \eta_b, \eta_{b+1} \right ], ~~
 \forall i = 1, 2, \ldots, k \right \}.
\ee
The definition of $\Sset$ is independent of $x^n$, and depends only on $\pvec$.
\begin{lemma}
\label{lemma_diverg_typical}
 Let $x^n \in \Tset_x$ and $\sigvec \in \Sset$.  Then,
 \be
  \label{eq:lem_diver_typ}
  \ln \frac{P_{\theta} \left (x^n \right )}{P_{\theta (\sigma)} \left ( x^n \right )} \leq
  \frac{6k}{n^{\varepsilon/2}} = o(k).
 \ee
\end{lemma}
Lemma~\ref{lemma_diverg_typical} shows that the probability of a typical $x^n$ given by a
permuted parameter vector in $\Sset$ diverges only by a negligible factor from
$P_{\theta} \left (x^n \right )$.  Its proof is in \ref{ap:lemma_diverg_typical_proof}.

Let $M_{\theta,\eta}$
be the number of permutation vectors $\sigvec$ in $\Sset$.
Using \eref{eq:lem_diver_typ}, for $x^n \in \Tset_x$,
\be
 \label{eq:ub1_p1}
 \log P_{\theta} \left [ \Psi \left ( x^n  \right ) \right ] \geq
 \log P_{\theta} \left ( x^n \right ) +
 \log M_{\theta,\eta} - o\left (k \right ),
\ee
\be
 \label{eq:Mtheta}
 M_{\theta,\eta} = \left | \Sset \right | = \prod_{b=2}^{B_{\eta}} k_b! =
 \prod_{b=2}^{A_{\eta}} k_b!.
\ee
Hence, applying \eref{eq:ub1_p1} and Lemma~\ref{lemma_untyp_prob} (step $(a)$ below),
\bea
 \nonumber
 H_{\theta} \left ( \Psi^n \right )
 &=&
 -\sum_{x^n \in \bar{\Tset}_x} P_{\theta} \left (x^n \right )
 \log P_{\theta} \left [\Psi \left (x^n \right ) \right ]
 -\sum_{x^n \in \Tset_x}
 P_{\theta} \left (x^n \right )
 \log P_{\theta} \left [\Psi \left (x^n \right ) \right ]
 \nonumber \\
 \nonumber \\
 &\stackrel{(a)}{\leq}&
 H_{\theta} \left ( X^n \right ) -
 \left ( 1 - \varepsilon_n \right )
 \log M_{\theta,\eta} + o(k)
 \nonumber \\
 &\leq&
 n H_{\theta} \left ( X \right ) -
 \left ( 1 - \varepsilon_n \right )
 \sum_{b=2}^{A_{\eta}} \log \left (k_b!\right ) + o(k).
 \label{eq:entropy_derivation_ub1}
\eea
The proof of Theorem~\ref{theorem:ub1} is concluded.
\end{proof}

\subsection{Lower Bounds}
\label{sec:large_lower}

\begin{theorem}
\label{theorem:lb1}
Fix $\delta > 0$.
Let $n \rightarrow \infty$ and $\varepsilon \geq (1+\delta)(\ln \ln n)/(\ln n)$.
If $\theta_i > 1/n^{1-\varepsilon}$, $\forall i, 1\leq i \leq k$,
\be
 \label{eq:lb1}
 H_{\theta} \left ( \Psi^n \right ) \geq
 n H_{\theta} \left ( X \right ) -
 \sum_{b=1}^{A_{\xi}} \log \left ( \kappa_b ! \right ) - k \log 3 - o(1),
\ee
and also
\be
 \label{eq:lb1a}
 H_{\theta} \left ( \Psi^n \right ) \geq
 n H_{\theta} \left ( X \right ) -
 \sum_{b=1}^{A_{\xi}} \log \left ( \kappa'_b ! \right ) - o(1).
\ee
\end{theorem}
The two bounds above are very close and except one step
are proved similarly.  The bound of \eref{eq:lb1} does not count occurrences
in a given bin more than once (except the correction term of $k \log 3$).
However, there exist distributions, such as the geometric distribution
(see, e.g., \cite{shamir07}), where components of $\pvec$ sparsely populate bins,
for which the bound of \eref{eq:lb1a} will be tighter.
The last term of $o(1)$ decays at an exponential rate of $O(\varepsilon_n n \log n)$,
where $\varepsilon_n$ is defined in \eref{eq:non_typ_prob}.
The pattern entropy is shown to decrease by logarithm
of the number of permutations within bins of $\xivec$.

\begin{proof}{}
Define the set of \emph{typical patterns\/} as
\be
 \label{eq:typical_pattern_lb1}
 \Tset_{\psi} \dfn
 \left \{ \psi^n ~:~ \exists x^n \in \Tset_x, \psi^n = \Psi \left ( x^n \right ) \right \}
\ee
the set of patterns, each of at least one typical sequence as defined in \eref{eq:typical_set}.
Now, for $\psi^n \in \Tset_{\psi}$,
let $M_{\theta, \xi} \left ( \psi^n \right ) \dfn \left |y^n \in \Tset_x ~:~
\psi^n = \Psi \left (y^n \right ) \right |$ be the number of typical sequences
that have the pattern $\psi^n$, and let $\bar{M}_{\theta, \xi}$ and
$\bar{M}'_{\theta, \xi}$
denote upper bounds
on $M_{\theta, \xi} \left ( \psi^n \right )$ for $\psi^n \in \Tset_{\psi}$.
\begin{lemma}
\label{lemma_typical_bound}
Let $\psi^n \in \Tset_{\psi}$.  Then,
\bea
 \label{eq:lemma_typ_bound1}
 M_{\theta, \xi} \left ( \psi^n \right ) &\leq& \bar{M}_{\theta, \xi} \dfn
 3^k \cdot \prod_{b=1}^{A_{\xi}}  \kappa_b !, \\
 \label{eq:lemma_typ_bound2}
 M_{\theta, \xi} \left ( \psi^n \right ) &\leq& \bar{M}'_{\theta, \xi} \dfn
 \prod_{b=1}^{A_{\xi}}  \kappa'_b !.
\eea
\end{lemma}
The proof of Lemma~\ref{lemma_typical_bound} is in \ref{ap:lemma_typical_bound_proof}.
It now follows that
\bea
 \nonumber
 H_{\theta} \left (\Psi^n \right )
 &=&
 H_{\theta} \left (X^n \right ) - H_{\theta} \left (X^n | \Psi^n \right ) \\
 \nonumber
 &\stackrel{(a)}{\geq}&
 H_{\theta} \left (X^n \right ) -
 P_{\theta} \left ( \Tset_x \right )
 H_{\theta} \left ( X^n | \Psi^n, \Tset_x \right ) -
 P_{\theta} \left ( \bar{\Tset}_x \right )
 H_{\theta} \left ( X^n | \Psi^n, \bar{\Tset}_x \right ) -
 h_2 \left [ P_{\theta} \left ( \Tset_x \right ) \right ] \\
 \nonumber
 &\stackrel{(b)}{\geq}&
 H_{\theta} \left (X^n \right ) -
 \log \bar{M}_{\theta, \xi} -
 \varepsilon_n \log k! - o\left (n \varepsilon_n \right ) \\
 &=&
 n H_{\theta} \left (X \right ) -
 \log \bar{M}_{\theta, \xi} - o(1)
 \label{eq:theorem_lb1_proof}
\eea
where $(a)$ follows from the chain rule, namely,
\bea
 H_{\theta} \left ( X^n~|~ \Psi^n \right )
 &=&
 H_{\theta} \left ( X^n, T ~|~ \Psi^n \right )
 ~=~
 H_{\theta} \left ( X^n ~|~ \Psi^n, T \right ) + H_{\theta} \left [ T ~|~ \Psi^n \right ] \\
 &=&
 \nonumber
 P_{\theta} \left ( \Tset_x \right )
 H_{\theta} \left ( X^n | \Psi^n, \Tset_x \right ) +
 P_{\theta} \left ( \bar{\Tset}_x \right )
 H_{\theta} \left ( X^n | \Psi^n, \bar{\Tset}_x \right ) +
 H_{\theta} \left [ T ~|~ \Psi^n \right ],
\eea
where $T$ is a Bernoulli random variable, taking value $1$ if $\Tset_x$ occurs,
and the last term of step $(a)$ of \eref{eq:theorem_lb1_proof} follows since conditioning reduces entropy.
Step $(b)$ of \eref{eq:theorem_lb1_proof} follows from $P_{\theta} \left ( \Tset_x \right ) \leq 1$,
$H_{\theta} \left ( X^n | \Psi^n, \Tset_x \right ) \leq \log \bar{M}_{\theta, \xi}$,
Lemma~\ref{lemma_untyp_prob},
$H_{\theta} \left ( X^n | \Psi^n, \bar{\Tset}_x \right ) \leq \log k!$,
and from $h_2 [P_{\theta} \left ( \Tset_x \right ) ] = o(n \varepsilon_n)$.
Substituting $\bar{M}_{\theta, \xi}$ from
\eref{eq:lemma_typ_bound1} in \eref{eq:theorem_lb1_proof} yields the bound of
\eref{eq:lb1}.  Similarly, using $\bar{M}'_{\theta, \xi}$ from
\eref{eq:lemma_typ_bound2} yields \eref{eq:lb1a}.
\end{proof}

\section{Bounds for Very Large Alphabets}
\label{sec:very}

The more general case is now considered, where there exist
alphabet letters with very small probabilities that may not occur in $x^n$.
Specifically, the effect of such letters on $H_{\theta} \left (\Psi^n \right )$ is considered.

\subsection{Upper Bounds}

General upper bounds are derived by designing a low-complexity (non-universal) sequential
probability assignment method for $\psi^n$, whose average description length serves as an upper
bound on $H_{\theta} \left (\Psi^n \right )$.   Instead of coding $\psi^n$ by itself,
the pair $\left (\psi^n, \beta^n \right )$ is jointly coded, where $\beta^n$ represents the
sequence of bins corresponding to letters in $x^n$.  Different grids
produce different bounds.  Examples and study of pattern entropy for specific
distributions in \cite{shamir07} demonstrate that tightness
depends on the specific source distribution.  One bound may be tighter for one
and another for another.
\begin{theorem}
\label{theorem:ub3}
Fix $\delta > 0$.
Let $n \rightarrow \infty$ and $\varepsilon \geq (1+\delta)(\ln \ln n)/(\ln n)$
(also for $\etavec$ in \eref{eq:eta_grid_def2}).  Then,
\bea
 \nonumber
 H_{\theta} \left ( \Psi^n \right ) &\leq&
 n H^{(0,1)}_{\theta} \left ( X \right ) -
 \left ( 1 - \varepsilon \right ) \sum_{b=2}^{A_{\eta}}
 \log \left ( k_b ! \right ) \\ &+&
 \nonumber
 \left ( n \varphi_1 - L_1 \right ) \log
 \left [ \min \left \{ k_1, n \right \} \right ] +
 n \varphi_1 h_2 \left ( \frac{L_1}{n\varphi_1} \right ) \\ &+&
 \label{eq:ub3}
 \left ( \frac{n^2}{2} \sum_{i=1}^{k_0} \theta_i^2 \right )
 \log  \left \{ \frac{2 e \cdot \varphi_0 \cdot \min \left \{k_0, n \right \}}
 {n \sum_{i=1}^{k_0} \theta_i^2} \right \}.
\eea
\end{theorem}
The bound in \eref{eq:ub3} consists of four major components: 1) the i.i.d.\ entropy
in which bins $0$ and $1$ of $\etavec$ are each packed into a point mass (the first term),
2) the gain in first occurrences of symbols $i$ with $\theta_i > 1/n^{1-\varepsilon}$
(the second term), 3) the loss in packing bin $1$ (the next
two terms), and 4) the loss in packing bin $0$ (the last term).
The sum of the third and fourth terms
in \eref{eq:ub3} decreases with $L_1$ for $k_1 \geq (1 +\varepsilon)n^{\varepsilon}$,
thus $L_1$ can be replaced by a lower bound as in
\eref{eq:mean_bin_bound}.

If the symbols in bins $0$ and $1$ formed by $\etavec$ are packed into a single point mass, a simpler
upper bound that uses $H_{\theta}^{(01)}$ and $\varphi_{01}$ instead of
$H_{\theta}^{(0,1)}$, and both
$\varphi_0$ and $\varphi_1$, respectively, can be obtained.  Using
$\tauvec$ instead of $\etavec$ produces other bounds.
\begin{corollary}
\label{theorem:ub3_cor2}
Fix $\delta > 0$.
Let $n \rightarrow \infty$ and $\varepsilon \geq (1+\delta)(\ln \ln n)/(\ln n)$
(also for $\etavec$ in \eref{eq:eta_grid_def2}).  Then,
\bea
 \nonumber
 H_{\theta} \left ( \Psi^n \right ) &\leq&
 n H^{(01)}_{\theta} \left ( X \right ) -
 \left ( 1 - \varepsilon \right ) \sum_{b=2}^{A_{\eta}}
 \log \left ( k_b ! \right ) + \\ & &
 \label{eq:ub3_c1}
 \left ( n \varphi_{01} - L_{01} \right ) \log
 \left [ \min \left \{ k_{01}, n \right \} \right ] +
 n \varphi_{01} h_2 \left ( \frac{L_{01}}{n\varphi_{01}} \right ).
\eea
Let $n \rightarrow \infty$ and $\varepsilon \geq 0$.  Then,
\bea
 \label{eq:ub3_c21}
 H_{\theta} \left ( \Psi^n \right ) &\leq&
  n H^{(0)}_{\theta} \left ( X \right ) +
  \left ( \frac{n^2}{2} \sum_{i=1}^{k_0} \theta_i^2 \right )
  \log \left ( \frac{2e \cdot \varphi_0 \cdot \min \left \{k_0, n \right \}}
  {n \sum_{i=1}^{k_0} \theta_i^2 }\right ),\\
 \nonumber
 H_{\theta} \left ( \Psi^n \right ) &\leq&
  n H^{(0)}_{\theta} \left ( X \right ) -
  \sum_{b=1}^{A_{\tau}} \sum_{m=0}^{c_b}
  P_{\theta} \left (C_b = m \right )
  \log \frac{c_b !}{\left ( c_b - m \right )!} +
  \frac{9 \log e}{n^{\varepsilon}} \sum_{b \geq 1, c_b > 1} c_b + \\ & &
 \label{eq:ub3_c2}
  \left ( \frac{n^2}{2} \sum_{i=1}^{k_0} \theta_i^2 \right )
  \log \left ( \frac{2e \cdot \varphi_0 \cdot \min \left \{k_0, n \right \}}
  {n \sum_{i=1}^{k_0} \theta_i^2 }\right ).
\eea
\end{corollary}
The bound in \eref{eq:ub3_c2} is in many cases the tightest but is harder to compute.
It can be simplified using Stirling's approximation,
\be
 \label{eq:stirling}
 \sqrt{2 \pi m} \left ( \frac{m}{e} \right )^m \leq m! \leq
 \sqrt{2 \pi m} \left ( \frac{m}{e} \right )^m \cdot e^{1/(12m)},
\ee
and Jensen's inequality,
at the expense of loosening it, by replacing the inner sum in its
second term by $\left ( E_{\theta} [C_b] \right ) \log \left \{\left ( E_{\theta} [C_b] \right )/e \right \}$,
where $E_{\theta} [C_b]$ is the expected distinct letter count in bin $b$ of $\tauvec$.
The bounds in \eref{eq:ub3} and \eref{eq:ub3_c1} trade off two costs: \eref{eq:ub3} has a larger
first term, while \eref{eq:ub3_c1} pays a higher penalty in its last two terms.
The better trade off is distribution dependent.
Roughly, if letters in bins $0$ and $1$ of $\etavec$ are better separated,
\eref{eq:ub3} is tighter, while otherwise \eref{eq:ub3_c1} is tighter.
The bound in \eref{eq:ub3_c21} is the simplest, but ignores gains of first occurrences
of letters with large probabilities.  Its best use is thus for fast decaying distributions.
Both \eref{eq:ub3_c21} and \eref{eq:ub3_c2} may be tightened in certain cases by separating
low probabilities (bin $0$ of $\etavec$) into two or more regions
(see, e.g., \cite{shamir07}).
The next examples illustrate tradeoffs between the bounds.

\begin{example}
\label{ex:uniform_k1}
For a uniform distribution with $k=k_1 = n^{1-\nu}$ parameters
$\theta_i = 1/n^{1-\nu}$, where $0 < \nu \leq \varepsilon$,
\be
 H_{\theta} \left ( \Psi^n \right ) \leq
 n \log n^{1-\nu} -
 n^{1-\nu} \log \frac{n^{1-2\nu}}{e}
 - \Theta \left ( n^{1-2\nu} \right )
\ee
with \eref{eq:ub3} and \eref{eq:ub3_c1}.
Bound \eref{eq:ub3_c21} produces only the first term.  Then, with the loosened \eref{eq:ub3_c2},
\be
 H_{\theta} \left ( \Psi^n \right ) \leq
 n \log n^{1-\nu} -
 n^{1-\nu} \log \frac{n^{1-\nu}}{e}
 + \Theta \left ( n^{1-\varepsilon-\nu} \right ).
\ee
The last bound from \eref{eq:ub3_c2} is the tightest.
\end{example}

\begin{example}
Let $\pvec$ consist of two sets of probabilities:
$k_0 = \varphi_0 n^{1+\mu}$; $\mu \geq \varepsilon$,
probabilities of $1/n^{1+\mu}$,
and $k_1 = \varphi_1 n^{1-\nu}$; $0 < \nu < \varepsilon$, probabilities of $1/n^{1-\nu}$,
where $\varphi_0 + \varphi_1 = 1$.  Then,
\be
 H_{\theta} \left ( \Psi^n \right ) \leq
 \left \{
 \begin{array}{ll}
 \left ( 1 - \nu \right ) n\varphi_1 \log n -
 n \varphi_0 \log \varphi_0 -
 \Theta \left ( n^{1-\nu} \log n \right ), &
 \mbox{using \eref{eq:ub3} and \eref{eq:ub3_c2}} \\
 n \varphi_1 \log n +
 nh_2 \left ( \varphi_0 \right ) -
 O \left ( n^{1-\nu} \log n \right ), &
 \mbox{using \eref{eq:ub3_c1}} \\
 \left ( 1 - \nu \right ) n\varphi_1 \log n -
 n \varphi_0 \log \varphi_0 +
 \Theta \left ( n^{1-\mu} \log n \right ), &
 \mbox{using \eref{eq:ub3_c21}.}
 \end{array}
 \right .
\ee
The bound from \eref{eq:ub3_c1} is looser
because it ignores the clear separation between bins $0$ and $1$.
The gain ignored in \eref{eq:ub3_c21} also slightly loosens
the resulting bound.  The greater $\varphi_0$ is, the smaller
$H_{\theta} \left ( \Psi^n \right )$ is from $nH_{\theta}(X)$.
\end{example}

\begin{example}
For a given $\varepsilon > 0$,
let $\pvec$ consist of two sets of probabilities:
$k_0 = \varphi_0 n^{1+\mu}$; $\mu \geq \varepsilon$,
probabilities of $1/n^{1+\mu}$,
and $k_1 = \varphi_1 n^{1+\nu}$; $0 < \nu < \varepsilon$, probabilities of $1/n^{1+\nu}$,
where $\varphi_0 + \varphi_1 = 1$.  Here, \eref{eq:ub3} results in
a bound of $nh_2\left ( \varphi_0 \right ) + O \left ( n^{1-\nu} \log n \right )$.
A much tighter bound of $\Theta \left ( n^{1-\nu} \log n \right )$ is produced by
\eref{eq:ub3_c1}.  This is because the two sets here are of ``small'' probabilities.
Looser bounds of $\Theta \left ( n \log n \right )$
are produced by \eref{eq:ub3_c21} and the loosened \eref{eq:ub3_c2}, with a smaller
coefficient for the second.
However, since $\varepsilon \geq 0$ for these two bounds, $\varepsilon < \nu$ can
be used to produce similar bounds to that of \eref{eq:ub3_c1}.  Such flexibility is
limited with the other bounds that have positive lower limits on $\varepsilon$.
\end{example}

\begin{example}
Let $\pvec$ consist of two sets of probabilities:
$k_0 = \varphi_0 n^{1+\mu}$; $\mu \geq \varepsilon$,
probabilities of $1/n^{1+\mu}$,
and $k_1 = \varphi_1 n$ probabilities of $1/n$,
where $\varphi_0 + \varphi_1 = 1$.  Then,
\be
 H_{\theta} \left ( \Psi^n \right ) \leq
 \left \{
 \begin{array}{ll}
 \frac{n \varphi_1}{e} \log n +
 n \left [h_2 \left ( \varphi_0 \right ) +
 \varphi_1 h_2 \left ( \frac{1}{e} \right ) +
 \frac{\varphi_1}{e} \log \varphi_1 \right ] +
 O \left ( n^{1-\mu} \log n \right ), &
 \mbox{with \eref{eq:ub3}} \\
 \frac{n \varphi_1}{e} \log n +
 n h_2 \left ( \frac{\varphi_1}{e} \right ) +
 O \left ( n^{1-\mu} \log n \right ), &
 \mbox{with \eref{eq:ub3_c1}} \\
 n \varphi_1 \log n - n \varphi_0 \log \varphi_0 +
 O \left ( n^{1-\mu} \log n \right ), &
 \mbox{with \eref{eq:ub3_c21}} \\
 \frac{n \varphi_1}{e} \log n +
 n \left [h_2 \left ( \varphi_0 \right ) +
 \frac{\varphi_1}{e} \log \frac{\varphi_1 \left ( 1 - \frac{1}{e} \right )}{e}  +
 \varphi_1 \log \frac{e}{1-\frac{1}{e}}
 \right ] +
 O \left ( n^{1-\mu} \log n \right ), &
 \mbox{with \eref{eq:ub3_c2}.}
 \end{array}
 \right .
\ee
Again, the tightest bound is from \eref{eq:ub3_c1}, implying that all probabilities here
are still ``small''.  The next is that of \eref{eq:ub3}.  Unlike the other examples,
\eref{eq:ub3_c2} and \eref{eq:ub3_c21} lead to the loosest bounds.
If \eref{eq:ub3_c2} is used with
$\varepsilon = 0$, an even weaker bound with first term
$0.5\varphi_1 n \log n$ will result because
of the use of the upper bound of
\eref{eq:mean_recur_bound0} for mean re-occurrence count, which is looser than that
of \eref{eq:mean_reoccur_bound}.  Using \eref{eq:mean_reoccur_bound}
instead for the last term of
\eref{eq:ub3_c21}-\eref{eq:ub3_c2}
yields the bound of \eref{eq:ub3_c1} for this case.
\end{example}

In Theorem~\ref{theorem:ub3}, Corollary~\ref{theorem:ub3_cor2}, and the examples above,
contributions of small probabilities influence the pattern entropy.  The next corollary
shows the limits of these contributions.
\begin{corollary}
\label{theorem:ub3_cor1}
I. The total combined contribution of all letters with $\theta \leq 1/n^{1-\varepsilon}$ to
$H_{\theta} \left ( \Psi^n \right )$ beyond the term $-n\varphi_{01} \log \varphi_{01}$ of
$n H_{\theta}^{(01)} (X)$ is upper bounded by the maximum between $O \left (n^{2\varepsilon} \log n \right )$ and
\[
 n \varphi_{01} \log \ell_{01}+
 \varphi_{01} n^{1-\varepsilon} \log \frac{e n^{\varepsilon}}{\ell_{01}} +
 \Theta \left ( \varphi_{01} n^{1-\varepsilon} e^{-n^{\varepsilon}}\right ) =
 O \left ( n \varphi_{01} \log n \right ),
\]
Similarly, the sum of the third and fourth term in \eref{eq:ub3} is
$O \left ( \max \left \{n \varphi_1, n^{2\varepsilon} \right \} \log n \right )$. \\
II. The total combined contribution of all letters with
$\theta_i \leq 1/n^{\mu + \varepsilon}$, for any $\mu \geq 1$, beyond the
term $-n\varphi_0 \log \varphi_0$ of $n H_{\theta}^{(0,1)} (X)$
is $O \left ( n^{2 - \mu - \varepsilon} \log n \right )$.
Similarly, the last term of \eref{eq:ub3} is upper bounded by
\[
 \frac{\varphi_0 \cdot n^{1-\varepsilon}}{2} \log \left ( 2 e n^{1+\varepsilon} \right ) =
 O \left ( n^{1-\varepsilon} \varphi_0 \log n \right ) = o \left ( n \right ).
\]
\end{corollary}
Corollary~\ref{theorem:ub3_cor1} is proved in \ref{ap:ub3_cor1_proof}.  It shows that
the per-symbol (normalized by $n$)
contribution of bin $0$ of $\etavec$ beyond a single point mass is diminishing.
Furthermore, any letter with $\theta_i \leq 1/n^{2+\varepsilon}$ has
diminishing contribution to the block entropy beyond that of the single point mass of bin $0$.
The subsection is concluded with the proof of Theorem~\ref{theorem:ub3} and Corollary~\ref{theorem:ub3_cor2}.

\begin{proof}{of Theorem~\ref{theorem:ub3} and Corollary~\ref{theorem:ub3_cor2}}
For some $x^n$, let $\psi^n = \Psi \left ( x^n \right )$, and define
$\beta^n = \left ( \beta_1, \beta_2, \ldots, \beta_n \right )$ by
$\beta_j = b$ if $\theta_{x_j} \in \left ( \eta_b, \eta_{b+1} \right ]$.
The joint sequence $\left ( \psi^n, \beta^n \right )$ is sequentially assigned
probability
\be
 \label{eq:seq_prob_total}
 Q \left [ \left (\psi^n, \beta^n \right ) \right ] \dfn
 \prod_{j=1}^n Q \left [ \psi_j, \beta_j ~|~ \left ( \psi^{j-1},
 \beta^{j-1} \right ) \right ],
\ee
where
\be
\label{eq:seq_prob}
  Q \left [ \psi_j, \beta_j ~|~ \left ( \psi^{j-1}, \beta^{j-1} \right ) \right ] =
  \left \{
  \begin{array}{ll}
   \rho_{\beta_j} &
   \mbox{if}~ \left ( \psi_j, \beta_j \right ) \in \left ( \psi^{j-1}, \beta^{j-1} \right ), \\
   \varphi_{\beta_j} - k_{\beta_j} \left [\left (\psi^{j-1},\beta^{j-1} \right ) \right ]
   \cdot \rho_{\beta_j} &
   \mbox{if}~ \psi_j = \max \left \{ \psi_1, \ldots, \psi_{j-1}  \right \} + 1, \\
   0 & \mbox{otherwise},
  \end{array}
  \right .
\ee
where $\rho_b \dfn \varphi_b / k_b$ for $b \geq 2$, $\rho_0$ and $\rho_1$
are optimized later, and
$k_{\beta_j} \left [\left (\psi^{j-1},\beta^{j-1} \right )\right ]$
is the number of distinct indices that jointly occurred with
bin index $\beta_j$ in $\left (\psi^{j-1}, \beta^{j-1} \right )$ (e.g., if
$\psi^{j-1} = 1232345$ and $\beta^{j-1} = 1222242$ then
$k_{\beta_j} \left [ \left (\psi^{7}, \beta^{7} \right ) \right ]$ is
$3$ for $\beta_j = 2$, $1$ for $\beta_j = 1$ and $\beta_j = 4$, and
is $0$ for any other value of $\beta_j$).
Initially, every
bin $b$ is assigned its total probability $\varphi_b$.  Each new index occurring with a letter in bin $b$ is
assigned the remaining probability in bin $b$ for its first occurrence.
For any re-occurrence, it is assigned the average symbol probability in $b$; $\rho_b$, unless
$b \leq 1$, where a different (smaller) value which favors first occurrences is used for $\rho_b$.
After a new occurrence of a symbol in bin $b$, $\rho_b$ is subtracted from the remaining
bin probability.

Since joint entropy is not smaller than the entropy of one of the components,
\bea
 \nonumber
 H_{\theta} \left ( \Psi^n \right ) &\leq&
 H_{\theta} \left ( \Psi^n, \Beta^n \right ) \leq
 -E \log Q \left ( \Psi^n, \Beta^n \right ) \\
 \nonumber
 &=&
 -\sum_{x^n \in \Sigma^n} P_{\theta} \left (x^n \right ) \sum_{j=1}^n
 \log Q \left \{ \Psi \left (x_j \right ), \beta \left (x_j \right ) ~|~
 \left [ \Psi \left ( x^{j-1} \right ), \beta \left (x^{j-1} \right ) \right ] \right \} \\
 \nonumber
 &\stackrel{(a)}{=}&
 -n \sum_{i=k_{01}+1}^k \theta_i \log \rho_b \left ( \theta_i \right ) -
 \sum_{b=2}^{B_{\eta}} \sum_{m=0}^{k_b} P_{\theta} \left ( K_b = m \right )
 \log \frac{k_b!}{\left (k_b - m \right )!} \\
 & &  -\sum_{b=0}^1 \underbrace{ \left \{ \left ( n \varphi_b - E_{\theta} \left [K_b \right ] \right )
 \log \rho_b +
 \sum_{m=0}^{\min \left \{ n, k_b \right \}}
 P_{\theta} \left (K_b = m \right )
 \sum_{l = 0}^{m-1} \log \left ( \varphi_b - l \rho_b \right ) \right \} }_{R_b},
\label{eq:ub3_description_length}
\eea
where $\rho_b \left ( \theta_i \right )$ is the mean symbol probability in
bin $b$, where $\theta_i \in \left ( \eta_b, \eta_{b+1} \right ]$.
Equality $(a)$ is obtained as follows:  The first term is the coding cost of
``large'' probability letters.  The second term describes the gain of first occurrences
of these letters.
The first symbol occurring in a bin is assigned probability $k_b \rho_b$ at first
occurrence, the second $(k_b - 1) \rho_b$, and so on.
The remaining terms $R_b$ describe similar costs for bins $0$ and $1$.
The first element for each is the re-occurrence cost.  The second
is the first occurrence cost.  Bounds on all terms are summarized below.

\begin{lemma}
\label{lemma_theorem_ub3_bounds}
\be
 \label{eq:ub3_term1}
 -n \sum_{i=k_{01}+1}^k \theta_i \log \rho_b \left ( \theta_i \right ) \leq
 n H_{\theta}^{(0,1)} (X) + n
 \sum_{b = 0}^1 \varphi_b \log \varphi_b  +
 \frac{9 \log e}{n^{2\varepsilon}} \cdot
 \sum_{b \geq 2, k_b > 1} k_b
\ee
\be
\label{eq:ub3_term2}
 - \sum_{b=2}^{B_{\eta}} \sum_{m=0}^{k_b} P_{\theta} \left ( K_b = m \right )
 \log \frac{k_b!}{\left (k_b - m \right )!} \leq
 -\sum_{b=2}^{A_{\eta}} \left [ 1 - k_b \exp
\left \{-n \eta_b \right \} \right ] \log \left (k_b! \right )
\ee
The optimal choice of $\rho_b$; $b=0,1$, is
\be
 \label{eq:rho_b_min}
 \rho_b = \frac{\left ( n \varphi_b - L_b \right ) \varphi_b }{n \varphi_b \ell_b} =
 \frac{\left ( n \varphi_b - L_b \right ) }{n \cdot
 \min \left \{k_b, n \right \}}.
\ee
With this choice,
\be
\label{eq:low_prob_cont1}
 R_b \leq -n\varphi_b \log \varphi_b +
 \left ( n \varphi_b - L_b \right ) \log \left [\min \left \{k_b, n \right \} \right ] +
 n \varphi_b \cdot h_2 \left ( \frac{L_b}{n\varphi_b} \right ), ~~b = 0, 1,
\ee
which decreases with $L_b$ for
$b = 0$ and also for $b=1$ if $k_1 \geq \left ( 1 + \varepsilon \right ) n^{\varepsilon}$.
Specifically,
\be
\label{eq:bin0_cont}
 R_0 \leq -n \varphi_0 \log \varphi_0 +
 \left ( \frac{n^2}{2} \sum_{i=1}^{k_0} \theta_i^2 \right )
 \log \frac{2 e \cdot \varphi_0 \cdot \min \left \{k_0, n \right \}}
 {n \sum_{i=1}^{k_0} \theta_i^2}.
\ee
\end{lemma}
Lemma~\ref{lemma_theorem_ub3_bounds} is proved in \ref{ap:lemma_theorem_ub3_bounds_proof}.
Summing \eref{eq:ub3_term1}, \eref{eq:ub3_term2}, \eref{eq:low_prob_cont1} for $b=1$,
and \eref{eq:bin0_cont} yields \eref{eq:ub3},
where the last term of \eref{eq:ub3_term1} and the decaying
terms in \eref{eq:ub3_term2}, which decay at least as fast as $k_b \exp \left \{ -n^{\varepsilon} \right \}$
each, are absorbed by the leading $\varepsilon$ of the second
term in \eref{eq:ub3}.  The decrease of \eref{eq:low_prob_cont1} with $L_1$ implies that
the lower bound on $L_1$ of \eref{eq:mean_bin_bound} can be used
in \eref{eq:ub3} as long as
$k_1 \geq (1+\varepsilon) n^{\varepsilon}$.

Corollary~\ref{theorem:ub3_cor2} follows from similar steps.  To
prove \eref{eq:ub3_c1}, bins $0$ and $1$ are grouped to one point mass.
Then, \eref{eq:ub3_term1} and \eref{eq:low_prob_cont1} are adjusted
with $H_{\theta}^{(01)}(X)$, $k_{01}$, $L_{01}$, and $\varphi_{01}$, and summed together
with \eref{eq:ub3_term2} to produce \eref{eq:ub3_c1}.
Bound \eref{eq:ub3_c21} is obtained by packing bin $0$ of $\tauvec$ into a point mass, but
coding each ``large'' probability symbol as an independent bin.  If, in addition,
the ``large'' probability bins of $\tauvec$ are coded as in proving
\eref{eq:ub3}, an additional gain as the left hand side of \eref{eq:ub3_term2} w.r.t.\
$\tauvec$ is achieved.  Using $\tauvec$, the denominator of the last
term of \eref{eq:ub3_term1} is $n^{\varepsilon}$ (as can be seen in
\eref{eq:ub3_term1_proof}).
\end{proof}

\subsection{Lower Bounds}

The main difficulty in deriving a general lower bound on $H_{\theta} \left ( \Psi^n \right )$
is separating between ``small'' probabilities $\theta_i \leq 1/n^{1-\varepsilon}$, whose symbols $i$
may or may not occur in $X^n$, and ``large'' probabilities, for which the results of Theorem~\ref{theorem:lb1}
can be used.  The key idea is to use an auxiliary Bernoulli indicator random sequence $Z^n$ to aid
in the separation.

\begin{theorem}
\label{theorem:lb2}
Fix $\delta > 0$.
Let $n \rightarrow \infty$ and $\varepsilon \geq (1+\delta)(\ln \ln n)/(\ln n)$, define
$\xivec$ with \eref{eq:xi_grid_def}.
Define $Z^n \dfn \left (Z_1, Z_2, \ldots, Z_n \right )$ by
$Z_j = 0$ if $\theta_{X_j} \leq 1/n^{1-\varepsilon}$, and $1$ otherwise.
Let $k_{\vartheta}^-$ be the count of letters $i$ such that
$\theta_i \in \left ( \vartheta^- / n^{1-\varepsilon}, 1/ n^{1-\varepsilon} \right ]$ and
$k_{\vartheta}^+$ the count of letters $i$ with
$\theta_i \in \left ( 1/ n^{1-\varepsilon}, \vartheta^+/ n^{1-\varepsilon} \right ]$,
where $\vartheta^-$, $\vartheta^+$ are constants that satisfy $\vartheta^+ > 1 > \vartheta^- > 0$.
Then,
\be
 \label{eq:lb2}
 H_{\theta} \left ( \Psi^n \right ) \geq
 n H^{(01)}_{\theta} \left ( X \right ) - S_1 + S_2 + S_3 - S_4 - o(1),
\ee
where
\bea
 \label{eq:lb2_S1b1}
 S_1 &\leq&
 \sum_{b=1}^{A_{\xi}} \log \left ( \kappa_b ! \right ) + \left (k - \kappa_0 \right ) \log 3, \\
 \label{eq:lb2_S1b2}
 S_1 &\leq& \sum_{b=1}^{A_{\xi}} \log \left ( \kappa'_b ! \right ),
\eea
\bea
 \label{eq:lb2_S2b0}
 S_2 &=&
 \sum_{i=1}^{k_{01}} E_{\theta}
 \left [ N_x(i) - P_{\theta} \left ( i \in X^n \right ) \right ]
 \log \frac{\varphi_{01}}{\theta_i}, \\
 \label{eq:lb2_S2b1}
 S_2 &\geq&
 \sum_{i=1}^{k_{01} - 1} \left [ n \theta_i - 1 + e^{-n \left ( \theta_i +
 \theta_i^2 \right )} \right ]
 \log \frac{\varphi_{01}}{\theta_i} +
 \left ( n \theta_{k_{01}} - 1 \right ) \log \frac{\varphi_{01}}{\theta_{k_{01}}},
\eea
\bea
 \label{eq:lb2_S2b2}
 S_2 &\geq&
 \left ( 1 - \frac{1}{n^{\varepsilon}} \right )
 \frac{n^2}{2} \sum_{i=1}^{k_0} \theta_i^2 \log \frac{\varphi_{01}}{\theta_i} +
 \sum_{i=k_0 + 1}^{k_{01} - 1} \left [ n \theta_i - 1 + e^{-n \left ( \theta_i +
 \theta_i^2 \right )} \right ]
 \log \frac{\varphi_{01}}{\theta_i},
\eea
\be
 \label{eq:lb2_S3b}
 S_3 \geq
 (\log e) \sum_{i=1}^{L_{01}-1} \left ( L_{01} - i \right )
 \frac{\theta_i}{\varphi_{01}},
\ee
\be
 \label{eq:lb2_S4}
 S_4 \leq
 \log \comb{k_{\vartheta}^- + k_{\vartheta}^+}{k_{\vartheta}^+} .
\ee
\end{theorem}
Theorem~\ref{theorem:lb2} lower bounds $H_{\theta} \left ( \Psi^n \right )$ in terms
of $H_{\theta}^{(01)} (X)$ with several correction terms, two of which are provided more than
one bound.
Term $S_1$ shows the decrease in $H_{\theta} \left ( \Psi^n \right )$ due to first
occurrences of letters $i$ with $\theta_i > 1/n^{1-\varepsilon}$.  Its bounds
are similar to the correction term in Theorem~\ref{theorem:lb1}.
Term $S_2$ is the cost of re-occurrences of letters with ``small'' probabilities.
Separation of the last term in \eref{eq:lb2_S2b1} from the sum is only necessary if
$\theta_{k_{01}} > 3/5$ (see \eref{eq:mean_reoccur_bound}
and discussion following it).  Equation
\eref{eq:lb2_S2b2} separates the sum of
\eref{eq:lb2_S2b0} into bins $0$ and $1$,
where the additional term of \eref{eq:lb2_S2b1} can be added to tighten the bound.
Term $S_3$ is the penalty in first occurrences
of ``small'' probability symbols beyond the single point mass they are packed to.
Its bound in \eref{eq:lb2_S3b} is obtained under a worst case assumption and may be tightened.
Term $S_4$ is the correction from separation to ``small'' and ``large'' probabilities.
The last term of $-o(1)$ absorbs all the lower order terms.
By proper equalities, \eref{eq:lb2} can be brought into several other forms including
forms in terms of $H^{(0)}_{\theta}(X)$ and $H^{(0,1)}_{\theta}(X)$.

\begin{proof}{}
Using $Z^n$,
\be
 \label{eq:z_entropy}
 H_{\theta} \left ( \Psi^n \right ) =
 H_{\theta} \left ( \Psi^n ~|~ Z^n \right ) +
 H_{\theta} \left ( Z^n \right ) -
 H_{\theta} \left ( Z^n ~|~ \Psi^n \right ).
\ee
By definition of $Z^n$,
\be
 \label{eq:z_seq_ent}
 H_{\theta} \left ( Z^n \right ) = n h_2 \left ( \varphi_{01} \right ) \dfn
 - \varphi_{01} n \log \varphi_{01} - \left (1 - \varphi_{01} \right )n
 \log \left ( 1 - \varphi_{01} \right ).
\ee
The third term is bounded in the following lemma, which is proved
in \ref{ap:lemma_low_prob_boundary_proof}.
\begin{lemma}
\label{lemma_low_prob_boundary}
\be
 \label{eq:z_cond_entropy}
 S_4 \dfn H_{\theta} \left ( Z^n ~|~ \Psi^n \right ) \leq
 \log \comb{k_{\vartheta}^- + k_{\vartheta}^+}{k_{\vartheta}^+} + o(1).
\ee
\end{lemma}
To bound the first term of \eref{eq:z_entropy}, define two new pattern sequences
$\dot{\psi}^n$ and $\ddot{\psi}^n$.  The first is defined as $\dot{\psi}_j = \phi$
if $z_j = 1$, and for the second $\ddot{\psi}_j = \phi$ if $z_j = 0$, where $\phi$ is
a \emph{do not care\/} symbol.  The other components of both
$\dot{\psi}^n$ and $\ddot{\psi}^n$ are the patterns of the remaining symbols in $x^n$,
respectively, i.e., $\dot{\psi}^n$ and $\ddot{\psi}^n$ are the patterns
of low and high probability symbols occurring in $x^n$, respectively.  In a similar manner,
define $\dot{x}^n$ and $\ddot{x}^n$, where $\dot{x}_j = \phi$ if $z_j = 1$, $\dot{x}_j = x_j$, otherwise,
and $\ddot{x}_j = \phi$ if $z_j = 0$, $\ddot{x}_j = x_j$, otherwise.
Now,
\be
 \label{eq:H_Psi_Z}
 H_{\theta} \left ( \Psi^n ~|~ Z^n \right ) =
 H_{\theta} \left (\ddot{\Psi}^n ~|~ Z^n \right ) +
 H_{\theta} \left (\dot{\Psi}^n ~|~ Z^n \right )
\ee
because up to deterministic labeling of pattern indices, the uncertainty
on both sides is equal.

Following the same steps in \eref{eq:theorem_lb1_proof},
\bea
 \nonumber
 H_{\theta} \left (\ddot{\Psi}^n ~|~ Z^n \right )
 &=&
 H_{\theta} \left (\ddot{X}^n ~|~ Z^n \right ) -
 E_{\theta} \left \{ P_{\theta} \left ( \Tset_x ~|~ Z^n \right )
 H_{\theta} \left ( \ddot{X}^n ~|~ \Psi^n, Z^n, \Tset_x \right ) ~|~ Z^n \right \} -  \\
 \nonumber
 & &
 E_{\theta} \left \{ P_{\theta} \left ( \bar{\Tset}_x ~|~ Z^n \right )
 H_{\theta} \left ( \ddot{X}^n ~|~ \Psi^n, Z^n, \bar{\Tset}_x \right ) ~|~ Z^n \right \} -
 H_{\theta} \left ( T~|~ \Psi^n, Z^n \right ) \\
 \label{eq:H_Psi_Z1}
 &\stackrel{(a)}{\geq}&
 \left( 1 - \varphi_{01} \right ) n H_{\theta} \left ( X ~|~ Z = 1 \right ) -
 S_1 - \varepsilon_n \log (k-\kappa_0)! - o\left (n \varepsilon_n \right )
\eea
where the external expectation is on $Z^n$, and $\Tset_x$ and $T$ are as defined
in Section~\ref{sec:large}.  Now, $S_1$ can be upper bounded
by either \eref{eq:lb2_S1b1} or \eref{eq:lb2_S1b2} following
bounds similar to \eref{eq:lemma_typ_bound1} and \eref{eq:lemma_typ_bound2}, respectively,
$H_{\theta} (X~|~Z=1)$ is the i.i.d.\ source entropy given only letters with
$\theta_i > 1/n^{1-\varepsilon}$ are drawn, and $(a)$
follows from $P_{\theta} \left ( \Tset_x ~|~ Z^n \right ) \leq 1$,
$H_{\theta} \left ( \ddot{X}^n | \Psi^n, Z^n, \Tset_x \right ) \leq S_1$,
$H_{\theta} \left ( \ddot{X}^n | \Psi^n, Z^n, \bar{\Tset}_x \right ) \leq \log (k-\kappa_0)!$, and then
$E_{\theta} \left \{ P_{\theta} \left ( \bar{\Tset}_x ~|~ Z^n \right ) ~|~ Z^n  \right \} =
E_{\theta} \left \{ P_{\theta} \left ( \bar{\Tset}_x ~|~ Z^n \right ) \right \} =
P_{\theta} \left ( \bar{\Tset}_x \right ) \leq \varepsilon_n$.
Finally, $H_{\theta} \left ( T~|~ \Psi^n, Z^n \right ) \leq
h_2 [P_{\theta} \left ( \Tset_x \right ) ] = o(n \varepsilon_n)$.
Summing
\eref{eq:z_seq_ent} and \eref{eq:H_Psi_Z1},
\be
 \label{eq:proof_lb2_combine}
 \left( 1 - \varphi_{01} \right ) n H_{\theta} \left ( X ~|~ Z = 1 \right ) - S_1 - o(1) +
 n h_2 \left ( \varphi_{01} \right ) =
 nH_{\theta}^{(01)} (X) - S_1 - o(1).
\ee

With the chain rule, and data processing,
\bea
 \nonumber
 H_{\theta} \left (\dot{\Psi}^n ~|~ Z^n \right )
 &=&
 \sum_{j=1}^n H_{\theta} \left ( \dot{\Psi}_j ~|~ \dot{\Psi}^{j-1}, Z^n \right )
 ~\geq~
 \sum_{j=1}^n H_{\theta} \left ( \dot{\Psi}_j ~|~ X^{j-1}, Z^n \right ) \\
 \nonumber
 &\stackrel{(a)}{=}&
- \sum_{i=1}^{k_{01}}
 E_{\theta} \left \{
 E_{\theta} \left [ N_x(i) - P_{\theta} \left ( i \in X^n \right ) ~|~ Z^n \right ] \right \}
 \log P_{\theta} \left (i ~|~ Z = 0 \right ) + S_3 \\
 &\stackrel{(b)}{=}&
 \underbrace{
 \sum_{i=1}^{k_{01}}
 E_{\theta} \left [ N_x(i) - P_{\theta} \left ( i \in X^n \right )\right ]
 \log \frac{\varphi_{01}}{\theta_i}}_{S_2} + S_3
 \label{eq:H_Psi_Z0}
\eea
where $S_2$ is the average cost of re-occurrence of letters $i$ with $\theta_i \leq 1/n^{1-\varepsilon}$,
and $S_3$ is the average cost of first occurrences of such letters. Step $(a)$ follows from
rearranging the sum into re-occurrences and first occurrences, where each is expressed over
all (small probability) alphabet symbols, $(b)$ follows from
$E \left \{ E \left [ U ~|~ V \right ] \right \} = E \left [ U \right ]$, for random variables
$U$ and $V$, and since
$-\log P_{\theta} \left (i ~|~ Z = 0 \right ) = \log \left ( \varphi_{01}/\theta_i \right )$.
This yields \eref{eq:lb2_S2b0}.  Then, \eref{eq:lb2_S2b1} and \eref{eq:lb2_S2b2}
follow from \eref{eq:mean_reoccur_bound} and \eref{eq:mean_recur_bound0}, respectively,
where the preceding $1/n^{\varepsilon}$ in the first sum
of \eref{eq:lb2_S2b2} follows
from the lower bound in \eref{eq:mean_recur_bound0}.  Now,
\bea
 \nonumber
 S_3
 &\stackrel{(a)}{\geq}&
 -E_{\theta} \left \{ E_{\theta} \left [
 \sum_{i=0}^{K_{01}-1} \log \left ( 1 - \sum_{j=1}^i \frac{\theta_i}{\varphi_{01}} \right )
 ~|~ Z^n \right ] \right \}
 ~\stackrel{(b)}{\geq}~
 (\log e)
 E_{\theta} \left \{ E_{\theta} \left [
 \sum_{i=0}^{K_{01}-1} \sum_{j=1}^i \frac{\theta_i}{\varphi_{01}}
 ~|~ Z^n \right ] \right \} \\
 &\stackrel{(c)}{\geq}&
 (\log e) \cdot
 \sum_{i=0}^{L_{01}-1} \sum_{j=1}^i \frac{\theta_i}{\varphi_{01}}
 ~\stackrel{(d)}{=}~
 (\log e) \cdot
 \sum_{i=0}^{L_{01}-1} \left ( L_{01} - i \right ) \frac{\theta_i}{\varphi_{01}},
 \label{eq:H_Psi_S3}
\eea
where $(a)$ follows because each new occurrence of an index is allocated the remaining
total probability, where in the worst case, letters occur in ascending order of probabilities,
$(b)$ follows from $-\log(1-x) \geq x\log e$, $(c)$ follows from Jensen's inequality, where the
function is convex in $K_{01}$ because
each increase in $K_{01}$ results in no smaller
increase of the expression than the previous increase in $K_{01}$.  Then,
$E_{\theta} K_{01} = L_{01}$ is used.  Finally, $(d)$ follows rearrangement of the double sum.
The proof is concluded by combining \eref{eq:proof_lb2_combine}, \eref{eq:H_Psi_Z0},
and \eref{eq:z_cond_entropy} to obtain all components of \eref{eq:z_entropy}, where the
bounds on all terms are provided in \eref{eq:lb2_S1b1}-\eref{eq:lb2_S4}.
\end{proof}

\section{Entropy Range}
\label{sec:large_range}

Bounds presented so far depend on the arrangement
of probability parameters in the probability space.  However, can we say
more than \eref{eq:entropy_bounds_simple} about the pattern entropy without
knowledge of this arrangement?  The answer is yes for large enough $k$ and sufficiently
large letter probabilities.
There are $O\left ( \sqrt{n^{1+\varepsilon}} \right )$ bins in $\tauvec$.
Due to the constraint $\sum \theta_i = 1$, very few of the larger parameter
bins are populated, essentially leading to $O \left ( n^{1/3 + \varepsilon} \right )$
populated bins.  If $k$ is greater, this forces more than a single letter
probability to populate a single bin, thus decreasing the pattern entropy.
The range of values the entropy can
take is bounded below.
\begin{theorem}
\label{theorem:range}
Fix $\delta > 0$.
Let $n \rightarrow \infty$, and $\varepsilon, \varepsilon_1 \geq (1+\delta)(\ln \ln k)/(\ln n)$.
Let $\theta_i > 1/n^{1-\varepsilon_1}$, $\forall i, 1\leq i \leq k$, and
let $k \geq n^{1/3+\varepsilon}$.  Then,
\be
 \label{eq:range_c}
  n H_{\theta} \left ( X \right ) - \log \left (k! \right ) \leq
   H_{\theta} \left ( \Psi^n \right )
  \leq
  n H_{\theta} \left ( X \right ) -
  \frac{3}{2} k \log \frac{k}{e n^{1/3+\varepsilon/2}}.
\ee
\end{theorem}
The bounds of Theorem~\ref{theorem:range}
give a range within which the pattern entropy must be.
For alphabets with $k \geq n^{1/3+\varepsilon}$, the entropy \emph{must\/} decrease essentially
by at least $1.5 \log \left (k / n^{1/3} \right )$ bits per
alphabet symbol.
All low order terms can be absorbed in the denominator $\varepsilon/2$ exponent.
Alternatively, a term of $O \left ( k \log \log n \right )$ can be included,
and the exponent is reduced to $\varepsilon/3$.
Asymptotically, $\varepsilon_1$ and $\varepsilon$ can be equal.  However, for practical
$n$, different values may be required to guarantee occurrence of
all letters, and that low order
terms do not overwhelm the decrease in entropy.
Figure~\ref{fig:entropy ratio}
demonstrates the region of decrease in the pattern entropy w.r.t.\
the i.i.d.\ one.  The upper region bound shown is the non-asymptotic one given
in \eref{eq:range_ub_nonasymp} when proving \eref{eq:range_c}.  Smaller order terms influence
the region for practical $n$.
The tightness of \eref{eq:range_c} depends on the particular source.
For uniform sources, the lower bound gives the true behavior.
For sources
with monotonic parameters,
the upper bound gives a more accurate behavior, as demonstrated in
the following example.

\bef
\begin{minipage}[c]{.50\textwidth}
\includegraphics
[bbllx=50pt,bblly=185pt,bburx=560pt,
bbury=598pt,height=7cm, width=8.5cm, clip=]{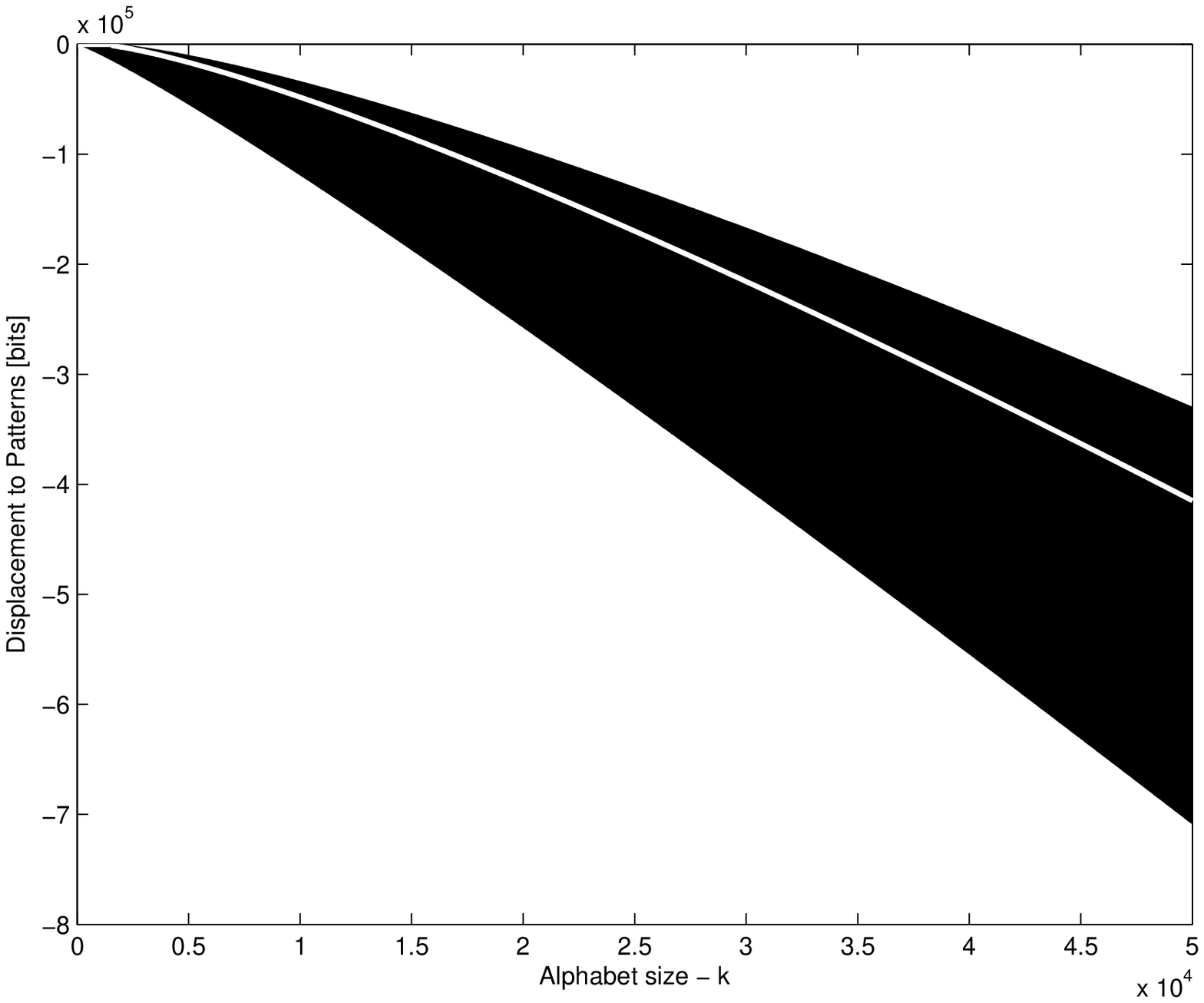}
\end{minipage}
\begin{minipage}[c]{.50\textwidth}
\includegraphics
[bbllx=50pt,bblly=185pt,bburx=560pt,
bbury=598pt,height=7cm, width=8.5cm, clip=]{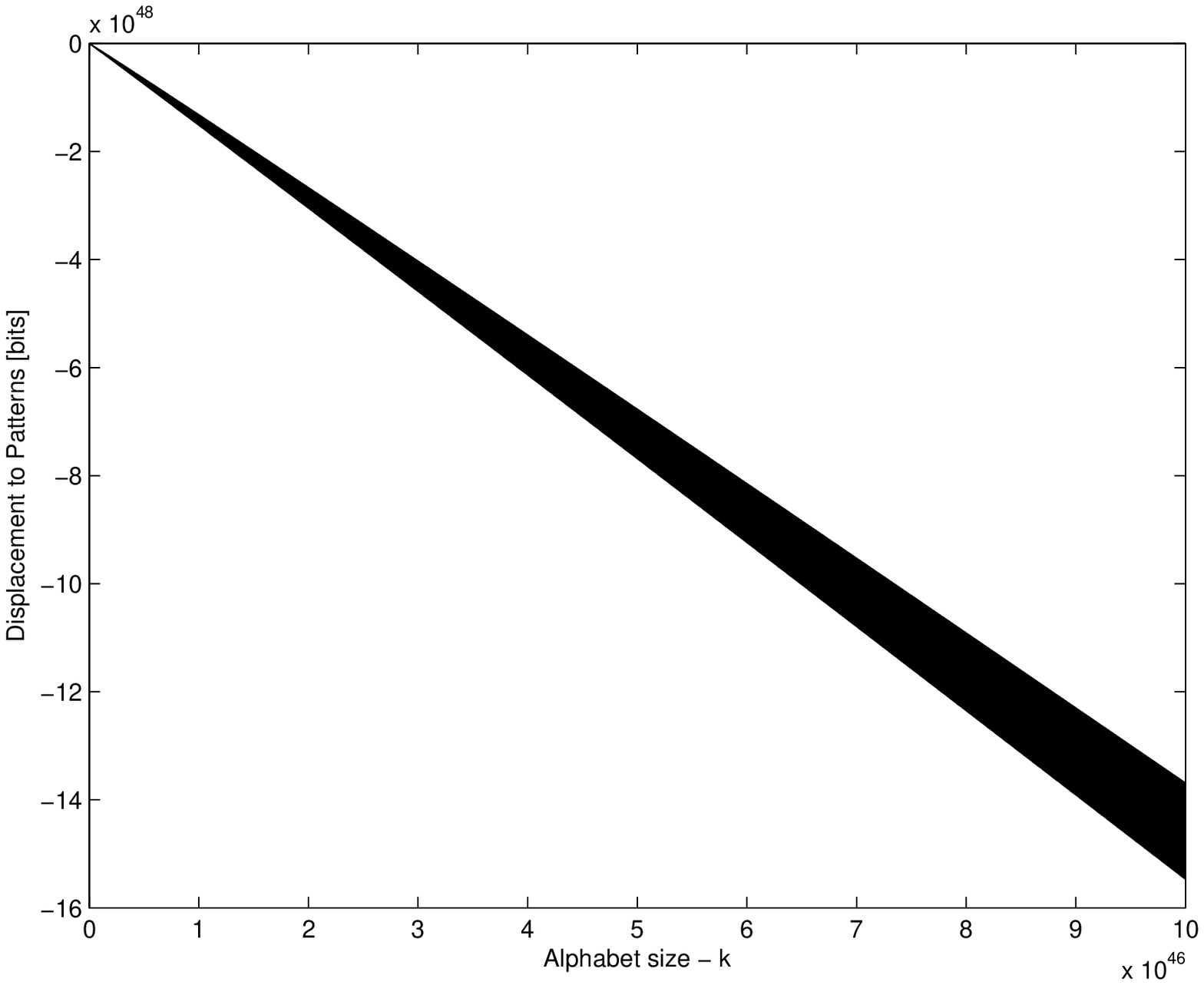}
\end{minipage}
\caption{Region of decrease from i.i.d.\ to pattern entropy vs.\
$k$ for $n=10^6$ bits, $\varepsilon = 0.2$, $n^{\varepsilon_1} = 20$ (left), and for
$n=10^{50}$ bits, $\varepsilon = 0.1$, $n^{\varepsilon_1} = 1000$ (right).  The solid white
curve on the left describes the asymptotic
decrease expression in \eref{eq:range_c}.  On the right, it overlaps the
non-asymptotic upper bound from \eref{eq:range_ub_nonasymp}.}
\label{fig:entropy ratio}
\enf

\begin{example}
Let $k = d \beta \geq n^{(1+\varepsilon)/3}$, where for
$b = 1, 2, \ldots, \beta$ there are $d$ letters with probability $\xi_b$.
Hence,
$\sum_i \theta_i = \sum_{b=1}^{\beta} d b^2/n^{1-\varepsilon} = 1$.
For $n^{1-\varepsilon} \gg k$, since $d\beta = k$,
$\beta \geq \sqrt{3n^{1-\varepsilon}/k} (1 - o(1))$.  This leads to
\be
 \label{eq:example_monotonic}
 \sum_{b=1}^{A_{\xi}} \log \left (\kappa_b! \right ) =
 \log \left [ \left ( \frac{k}{\beta} \right )! \right ]^{\beta} \leq
 \left ( 1 + o(1) \right ) k \log \frac{k}{\beta e} =
 \left ( 1 + o(1) \right )
 \frac{3}{2} k \log \frac{k}{e^{2/3} n^{(1-\varepsilon)/3} 3^{1/3}}.
\ee
Substituting  \eref{eq:example_monotonic} in \eref{eq:lb1} (with the third term of
\eref{eq:lb1} omitted because the letters in adjacent bins are sufficiently spaced),
the resulting lower bound asymptotically achieves
the upper limit in \eref{eq:range_c}.
\end{example}

\begin{proof}{of Theorem~\ref{theorem:range}}
The upper bound is proved by deriving an upper bound on the second
term of \eref{eq:ub3_c2} in Corollary~\ref{theorem:ub3_cor2}, which is determined by a lower bound
on $M_{\theta, \tau}$,
following a similar bound w.r.t.\ $\tauvec$ to that in
\eref{eq:ub3_term2}.  Since $\theta_i > 1/n^{1-\varepsilon_1} > 1/n^{1+\varepsilon}$,
only the first three terms of \eref{eq:ub3_c2} exist.  The first equals $nH_{\theta}(X)$ because
$k_0 = 0$.  Now,
for an arbitrary $\beta < B_{\tau}$, the set of bins formed by
$\tauvec$ is partitioned into two parts: all bins up to $\beta$
and all others.  The maximal possible number of components of $\pvec$ is allocated
to the second group, and the remaining components
are distributed in the first
group so that $M_{\theta, \tau}$ is minimized.
Then, since this holds for every $\beta$, $\beta$ that maximizes the lower bound
on $M_{\theta,\tau}$ is chosen.

For convenience, denote $A \dfn n^{1+ \varepsilon}$.
For $\theta_i \in \left (\tau_{\beta}, \tau_{\beta+1} \right ]$,
$\theta_i > \tau_{\beta}$.
From \eref{eq:tau_grid_def} and since $\sum_i \theta_i = 1$, it follows that
$\sum_{b=\beta}^{B_{\tau}} c_b = k - \sum_{b=1}^{\beta-1} c_b < A/{\beta}^2$.  Hence,
\be
 \label{eq:bbeta}
 \sum_{b=1}^{\beta-1} c_b > k - \frac{A}{\beta^2}.
\ee
An infimum on $M_{\theta, \tau}$ is obtained by uniformly distributing $k-A/\beta^2$ symbol
probabilities in $\beta$ bins, where the remaining symbol probabilities are uniformly placed
in all bins of $\tauvec$ (this is a lower bound because it may violate $\sum_i \theta_i = 1$).
Following an equation similar to \eref{eq:Mtheta} w.r.t.\ $\tauvec$,
\be
\label{eq:M_bound1}
 M_{\theta,\tau} \geq
 \left [ \left ( \frac{k - A/\beta^2}{\beta}\right )!\right ]^{\beta}
\ee
for every $\beta$. Applying Stirling's approximation \eref{eq:stirling},
\be
\label{eq:lnM_bound}
 \ln M_{\theta, \tau} \geq
 \left ( k - \frac{A}{\beta^2} \right )
 \ln \frac{k - A/\beta^2}{e\beta} +
 \frac{\beta}{2} \ln \frac{2 \pi \left ( k - A/\beta^2 \right )}{\beta}.
\ee
By differentiation, \eref{eq:lnM_bound} is shown to be maximized by
$\beta = \sqrt{\gamma A/k}$, where $\gamma \geq 2$ satisfies
\be
 \label{eq:gamma_max}
 \gamma = \ln \frac{(\gamma-1)^2}{\gamma^3} + \ln \frac{ek^3}{A}.
\ee
For $k \geq n^{1/3 +\varepsilon}$, this implies that $\gamma$ must
increase at $O( \ln n)$.  Thus, to first order,
\be
 \label{eq:b_opt}
 \beta_{opt} = \sqrt{\frac{\alpha A}{k} \ln \frac{k^3}{A}},
\ee
where $\alpha$ is a constant, asymptotically optimized slightly below $\alpha = 1$.
(The exact value of $\alpha$ only affects second order terms.)
Plugging \eref{eq:b_opt} with $\alpha = 1$ in \eref{eq:lnM_bound},
\be
 \label{eq:lnM_bound1}
 \ln M_{\theta, \tau} \geq
 \frac{3}{2} k \ln \frac{k}
 {e n^{1/3 + \varepsilon/3}} -
 \frac{k}{2} \left ( 1-\frac{1}{\ln \frac{k^3}{n^{1+\varepsilon}}} \right )
 \ln \ln \frac {k^3}{n^{1+\varepsilon}},
\ee
as long as $k \geq n^{1/3 + \varepsilon}$.
Plugging \eref{eq:lnM_bound1} in the second term of \eref{eq:ub3_c2},
using the upper bound of \eref{eq:no_letter_i_bound} on the probability of no occurrence
of any letter,
\be
 \label{eq:range_ub_nonasymp}
 H_{\theta} \left ( \Psi^n \right ) \leq nH_{\theta}(X) -
 \left ( 1 - ke^{-n^{\varepsilon_1}} \right )
 \left [ \frac{3}{2} k \log \frac{k}
 {e n^{1/3 + \varepsilon/3}} -
 \frac{k}{2} \left ( 1-\frac{1}{\ln \frac{k^3}{n^{1+\varepsilon}}} \right )
 \log \ln \frac {k^3}{n^{1+\varepsilon}} \right ] +
 \frac{9 k \log e}{n^{\varepsilon}}.
\ee
With the valid choices of $\varepsilon$ and $\varepsilon_1$, all lower order terms can be
absorbed in a term of $0.25\varepsilon k \log n$ for some sufficiently large $n$,
and the upper bound of
\eref{eq:range_c} follows.
\end{proof}

\section{Summary and Conclusions}

The entropy of patterns of i.i.d.\ sequences was studied.
Tight upper and lower bounds as function
of an i.i.d.\ source entropy, the alphabet size, the letter probabilities,
and their arrangement
in the probability space were derived first for distributions with bounded
probabilities, and then for the general case.
The bounds demonstrated
the range of values the pattern entropy can take, and showed that in many cases
it \emph{must\/} decrease substantially from the original i.i.d.\ sequence entropy.
It was shown that low probability symbols contribute mostly as a single
point mass to the pattern entropy.  However, an additional correction term is
necessary.  Very low probability
symbols contribute negligibly over the contribution of a single point mass.
The bounds obtained can be used to provide very accurate approximations of
the pattern block entropies for various distributions as shown in a followup paper
\cite{shamir07}.

\appendix
\renewcommand{\thesection}{Appendix \Alph{section}}
\renewcommand{\theequation}{\thesection.\arabic{equation}}

\section{--~~ Proof of Lemma~\ref{lemma_untyp_prob}}
\label{ap:lemma_untyp_prob_proof}
\renewcommand{\theequation}{A.\arabic{equation}}
\renewcommand{\theproposition}{A.\arabic{proposition}}
\renewcommand{\thelemma}{A.\arabic{lemma}}
\setcounter{equation}{0}
\setcounter{lemma}{0}
\setcounter{proposition}{0}

The set $\bar{\Tset}_x = \bigcup_i \eventA_i$, where
\be
\label{eq:event_Ai_def}
 \eventA_i = \left \{ x^n ~:~
  \left |\hat{\theta}_i - \theta_i \right | \geq
  \frac{\sqrt{\theta_i}}{2\sqrt{n}^{1-\varepsilon}}
 \right \}.
\ee
Using large deviations analysis of typical sets \cite{cover91}, \cite{cizar81},
\be
 \label{eq:event_Ai_prob}
 P_{\theta} \left ( \eventA_i \right ) \leq
 n \cdot 2^{-n \min_{\eventA_i} D \left ( \hat{\theta}_i || \theta_i \right )}
 \leq
 n \cdot 2^{-n \min \left [ D \left ( \theta_i + d_i || \theta_i \right ),
 D \left ( \theta_i - d_i || \theta_i \right ) \right ]}
\ee
where $d_i \dfn \sqrt{\theta_i}/(2\sqrt{n}^{1-\varepsilon})$ and
$D \left (\hat{\theta}_i || \theta_i \right )$ is the
\emph{divergence\/} (relative entropy) between the two Bernoulli distributions given
by $\hat{\theta}_i$ and $\theta_i$, respectively.  The coefficient $n$ is a bound on the number of types.
Using Taylor series expansions, for $n^{-(1-\varepsilon)} < \theta_i \leq 0.5$,
\bea
 \nonumber
 D \left ( \theta_i \pm d_i  ||\theta_i \right )
 &=&
 \left ( \theta_i \pm d_i \right )
 \log \left (1 \pm \frac{d_i}{\theta_i} \right ) +
 \left ( 1 - \theta_i \mp d_i \right )
 \log \left ( 1 \mp \frac{d_i}{1 - \theta_i} \right ) \\
 \nonumber
 &\stackrel{(a)}{\geq}&
 \log e \cdot \left \{
 \frac{d_i^2}{2\theta_i}
 \left ( 1 \mp \frac{d_i}{3\theta_i} \right ) +
 \frac{d_i^2}{2\left ( 1 -\theta_i \right )}
 \left ( 1 \pm \frac{d_i}{3 \left ( 1 - \theta_i \right )}
 \right ) \right \} \\
 \label{eq:diver_bound_untyp}
 &\stackrel{(b)}{\geq}&
 \frac{5 \log e}{48 n^{1-\varepsilon}} >
 \frac{\log e}{10 n^{1-\varepsilon}}
\eea
where $\pm$ and $\mp$ are used respectively to compactly describe both cases.
Step $(a)$ is obtained by combining the first three terms of the expansions for each of
the two logarithmic expressions.  The first terms from both expansions cancel each other.  Under
the assumptions bounding $\theta_i$, the remaining terms are all nonnegative, yielding a lower bound.
Plugging in the value of $d_i$, bounding $\theta_i$, the second term is now nonnegative negligible.
For the worst case, $1 - d_i/(3\theta_i) \geq 5/6$, leading to $(b)$.
Using the relation between divergence and $L_1$ distance (see, e.g., \cite{cover91}), for $\theta_i > 0.5$,
\be
 \label{eq:diver_bound_untyp_large}
 D \left ( \theta_i \pm d_i || \theta_i \right )
 \geq
 \frac{1}{2\ln 2}
 \left \| \left ( \theta_i \pm d_i \right ) - \theta_i \right \|_1^2
 = 2 (\log e) d_i^2
 \geq \frac{\log e}{4n^{1-\varepsilon}} >
 \frac{\log e}{10 n^{1-\varepsilon}}
\ee
where $\left \| \left ( \theta_i \pm d_i \right ) - \theta_i \right \|_1$ is the $L_1$ distance
between the Bernoulli distributions defined by $\theta_i \pm d_i$ and $\theta_i$, respectively.
Applying the union bound on the bounds in \eref{eq:diver_bound_untyp} and \eref{eq:diver_bound_untyp_large}
plugged in \eref{eq:event_Ai_prob},
\be
 \label{eq:proof_untyp_ub}
 P_{\theta} \left ( \bar{\Tset}_x \right ) \leq
 k \cdot n \cdot 2^{-0.1 (\log e) n^{\varepsilon}} \leq
 \exp \left \{-0.1 n^{\varepsilon} + (2 - \varepsilon) \ln n \right \},
\ee
where the second inequality follows from the limit on $\theta_i$ implying $k \leq n^{1-\varepsilon}$.
The bound is meaningful for $\varepsilon > (\ln \ln n + \ln 20)/(\ln n)$ and diminishes
for $\varepsilon \geq (1+\delta) (\ln \ln n)/(\ln n)$.
\hfill $\Box$

\section{--~~ Proof of Lemma~\ref{lemma_diverg_typical}}
\label{ap:lemma_diverg_typical_proof}
\renewcommand{\theequation}{B.\arabic{equation}}
\renewcommand{\theproposition}{B.\arabic{proposition}}
\renewcommand{\thelemma}{B.\arabic{lemma}}
\setcounter{equation}{0}
\setcounter{lemma}{0}
\setcounter{proposition}{0}

For a source $\pvec$, a permutation vector $\sigvec$, and a sequence $x^n$, define
\bea
 \delta_i &\dfn& \theta_i - \theta(\sigma_i) \\
 \hat{\delta}_i &\dfn& \hat{\theta}_i - \theta(\sigma_i).
\eea
Then, by the conditions of the lemma, the definition of $\Sset$ in \eref{eq:Sset_def},
and by \eref{eq:tau_grid_spacing},
\be
 \label{eq:good_permut}
 \left | \delta_i \right | \leq
 \frac{3 \sqrt{\theta(\sigma_i)}}{\sqrt{n}^{1+2\varepsilon}}.
\ee
By the triangle inequality,
\bea
 \left | \hat{\delta}_i \right | &=&
 \left | \hat{\theta}_i - \theta \left ( \sigma_i \right ) \right | ~\leq~
 \left | \hat{\theta}_i - \theta_i \right | +
 \left | \theta_i - \theta \left ( \sigma_i \right ) \right |
 \nonumber \\
 &\stackrel{(a)}{\leq}&
 \frac{\sqrt{\theta_i}}{2\sqrt{n}^{1-\varepsilon}} +
 \frac{3 \sqrt{\theta \left ( \sigma_i \right )}}
 {\sqrt{n}^{1 + 2\varepsilon}}
 ~=~
 \frac{\sqrt{\theta \left (\sigma_i \right ) + \delta_i}}{2\sqrt{n}^{1-\varepsilon}} +
 \frac{3 \sqrt{\theta \left ( \sigma_i \right )}}
 {\sqrt{n}^{1 + 2\varepsilon}}
 \nonumber \\
 &\stackrel{(b)}{\leq}&
 \frac{\sqrt{\theta \left (\sigma_i \right )}}{2\sqrt{n}^{1-\varepsilon}} +
 \frac{\sqrt{\left |\delta_i \right |}}{2\sqrt{n}^{1-\varepsilon}} +
 \frac{3 \sqrt{\theta \left ( \sigma_i \right )}}
 {\sqrt{n}^{1 + 2\varepsilon}}
 \nonumber \\
 &\stackrel{(c)}{\leq}&
 \frac{ \sqrt{\theta \left (\sigma_i \right )}}{\sqrt{n}^{1-\varepsilon}} +
 \frac{\sqrt{3} \theta \left ( \sigma_i \right )^{1/4}}{2n^{3/4}}
 ~\stackrel{(d)}{\leq}~
 \frac{2 \sqrt{\theta \left (\sigma_i \right )}}{\sqrt{n}^{1-\varepsilon}},
 \label{eq:delta_typ_bound1}
\eea
$(a)$ is obtained from \eref{eq:typical_set} and \eref{eq:good_permut},
$(b)$ is since $\sqrt{a+b} \leq \sqrt{a} + \sqrt{b}$ for $a,b \geq 0$, $(c)$
is from combining the first and last term and from \eref{eq:good_permut},
and $(d)$ results from
$\theta_i > 1/n^{1-\varepsilon} \Rightarrow 1/n^{1/4}
< \theta\left (\sigma_i \right )^{1/4}$.

Applying \eref{eq:good_permut}-\eref{eq:delta_typ_bound1}, for $x^n \in \Tset_x$ and $\sigvec \in \Sset$,
\bea
 \nonumber
 \ln \frac{P_{\theta} \left (x^n \right )}
 {P_{\theta(\sigma)} \left (x^n \right )}
 &=&
 \sum_{i=1}^k n \hat{\theta}_i \ln
 \frac{\theta_i}{\theta(\sigma_i)}
 ~=~
 n \sum_{i=1}^k \hat{\theta}_i \ln
 \left ( 1 + \frac{\delta_i}{\theta(\sigma_i)} \right )
 \\
 \nonumber
 &\stackrel{(a)}{\leq}&
 n \sum_{i=1}^k \hat{\theta}_i \frac{\delta_i}{\theta(\sigma_i)}
 ~\stackrel{(b)}{=}~ n \sum_{i=1}^k \left [ \delta_i +
 \frac{\hat{\delta}_i \delta_i}{\theta \left ( \sigma_i \right )} \right ] \\
 &\stackrel{(c)}{=}&
 n \sum_{i=1}^k \frac{\hat{\delta}_i \delta_i}{\theta(\sigma_i)}
 ~\stackrel{(d)}{\leq}~
 \frac{6k}{n^{\varepsilon/2}} = o(k)
 \label{eq:ratio_bound}
\eea
where $(a)$ follows from $\ln(1+x) \leq x$, $(b)$
follows from
$\hat{\theta}_i = \theta \left ( \sigma_i \right ) + \hat{\delta}_i$.
$(c)$ is because all displacements sum to $0$,
and $(d)$ follows by applying \eref{eq:good_permut}-\eref{eq:delta_typ_bound1}.
\hfill $\Box$

\section{--~~ Proof of Lemma~\ref{lemma_typical_bound}}
\label{ap:lemma_typical_bound_proof}
\renewcommand{\theequation}{C.\arabic{equation}}
\renewcommand{\theproposition}{C.\arabic{proposition}}
\renewcommand{\thelemma}{C.\arabic{lemma}}
\setcounter{equation}{0}
\setcounter{lemma}{0}
\setcounter{proposition}{0}

Let $\sigvec = \left \{ \sigma_j \right \}_{j=1}^k$ be a permutation vector.
Then, $x^n$ is permuted by $\sigvec$ to $w^n = \sigma \left (x^n \right ) \dfn
\left ( \sigma_{x_1}, \sigma_{x_2}, \ldots, \sigma_{x_n} \right )$.
For example, let $x^n = 333112222222$ and $\sigvec = \left (3, 1, 2 \right )$,
then, $w^n = 22233111111$, i.e., if $\sigma_j = i$, letter $j$ in $x^n$ is replaced
by $i$ in $w^n$.  In the example, $\sigma_2 = 1$, and $j=2$ is replaced by $i=1$.
We show that if $x^n, w^n \in \Tset_x$, then letter $i$ can replace only letters $j$
whose probability parameters are in the same bin as $\theta_i$ or in the two surrounding
bins of $\xivec$.  Then, the total number of such permutation vectors is upper bounded.

\begin{lemma}
\label{lemma_typ_permut}
 Let $\psi^n \in \Tset_{\psi}$, and $x^n \in \Tset_x$ such that
 $\psi^n = \Psi \left (x^n \right )$.  Let $w^n = \sigma \left (x^n \right )$
 such that $w^n \in \Tset_x$.  For $i; 1 \leq i \leq k,$ let
 $\theta_i \in \left (\xi_b, \xi_{b+1} \right ]$ and let $j$ be such that
 $\sigma_j = i$.  Then,
 $\theta_j \in \left ( \xi_{b-1}, \xi_{b+2} \right ]$.
\end{lemma}
\begin{proof}{}
The proof is by contradiction, $\theta_j \not \in
 \left ( \xi_{b-1}, \xi_{b+2} \right ]$ contradicts $x^n, w^n \in \Tset_x$.
By  $x^n, w^n \in \Tset_x$,
\bea
 \label{eq:lem_typ_p1}
 \left | \hat{\theta}_j \left ( x^n \right ) - \theta_j \right | &<&
 \frac{\sqrt{\theta_j}}{2 \sqrt{n}^{1-\varepsilon}}, \\
 \label{eq:lem_typ_p2}
 \left | \hat{\theta}_i \left (w^n \right ) - \theta_i \right | &<&
 \frac{\sqrt{\theta_i}}{2\sqrt{n}^{1-\varepsilon}}
\eea
where $\hat{\theta}_j \left ( x^n \right )$ and $\hat{\theta}_i \left ( w^n \right )$
are the ML estimates of $\theta_j$ and $\theta_i$ from $x^n$ and $w^n$, respectively.
By definition of $j$, $\hat{\theta}_i \left (w^n \right ) = \hat{\theta}_j \left ( x^n \right )$.
By the triangle inequality, \eref{eq:lem_typ_p1}, and \eref{eq:lem_typ_p2},
\bea
 \left | \theta_i - \theta_j \right |
 &\leq&
 \left |\theta_i - \hat{\theta}_j \left ( x^n \right ) \right | +
 \left | \hat{\theta}_j \left (x^n \right ) - \theta_j \right |
 \nonumber \\
 &=&
 \left |\theta_i - \hat{\theta}_i \left ( w^n \right ) \right | +
 \left | \hat{\theta}_j \left (x^n \right ) - \theta_j \right |
 \nonumber \\
 &<&
 \frac{\sqrt{\theta_i}}{2\sqrt{n}^{1-\varepsilon}} +
 \frac{\sqrt{\theta_j}}{2\sqrt{n}^{1-\varepsilon}} =
 \frac{\sqrt{\theta_i}+\sqrt{\theta_j}}{2\sqrt{n}^{1-\varepsilon}}.
 \label{eq:lem_typ_p3}
\eea
If $\theta_j \not \in \left ( \xi_{b-1}, \xi_{b+2} \right ]$, it must satisfy
$\theta_j \in \left ( \xi_{\beta}, \xi_{\beta+1} \right ]$ for some
$\beta \geq b+2$ or $\beta \leq b-2$.  In the first case,
\bea
 \theta_j - \theta_i &\geq& \xi_{\beta} - \xi_{\beta-1}
 \stackrel{(a)}{=}
 \frac{2 \left ( \beta + 1 - 1.5 \right )}{n^{1-\varepsilon}}
 \stackrel{(b)}{=}
 \frac{2\sqrt{\xi_{\beta+1}} \sqrt{n}^{1-\varepsilon} - 3}
 {n^{1-\varepsilon}}
 \nonumber \\
 &\stackrel{(c)}{\geq}&
 \frac{\sqrt{\xi_{\beta+1}}}{\sqrt{n}^{1-\varepsilon}}
 \stackrel{(d)}{\geq}
 \frac{\sqrt{\theta_j}}{\sqrt{n}^{1-\varepsilon}}
 \stackrel{(e)}{\geq}
 \frac{\sqrt{\theta_j} + \sqrt{\theta_i}}{2\sqrt{n}^{1-\varepsilon}},
 \label{eq:lem_typ_p4}
\eea
$(a)$ follows from \eref{eq:xi_grid_spacing} with $b = \beta-1$,
$(b)$ from \eref{eq:xi_grid_def} with $b = \beta+1$,
$(c)$ is because $\beta+1 \geq b+3 \geq 4$ and thus
$\sqrt{\xi_{\beta+1}} \geq 4/\sqrt{n}^{1-\varepsilon}$,  $(d)$ is by the
assumed range of $\theta_j$, and $(e)$ is again by the assumption that
$\theta_j > \theta_i$.
For $\theta_i > \theta_j$, where $\beta \leq b-2$, in a similar manner,
\be
 \label{eq:lem_typ_p5}
 \theta_i -\theta_j \geq
 \xi_b - \xi_{b-1} =
 \frac{2 \left ( b + 1 - 1.5 \right )}{n^{1-\varepsilon}} \geq
 \frac{\sqrt{\theta_j} + \sqrt{\theta_i}}{2\sqrt{n}^{1-\varepsilon}}.
\ee
The last inequality is obtained as \eref{eq:lem_typ_p4} by exchanging the roles
of $b$ and $\beta$.  Equations \eref{eq:lem_typ_p4} and \eref{eq:lem_typ_p5} contradict
\eref{eq:lem_typ_p3}.  Hence, $\theta_j \in \left ( \xi_{b-1}, \xi_{b+2} \right ]$.
\end{proof}

Equation \eref{eq:lemma_typ_bound2} follows directly from Lemma~\ref{lemma_typ_permut}, since
every letter $i$ can only permute into $\left ( \xi_{b-1}, \xi_{b+2} \right ]$.
To prove \eref{eq:lemma_typ_bound1}, a permutation replacing letter $j$
with $\theta_j \in \left ( \xi_{b-1}, \xi_{b+2} \right ]$ by letter $i$
with $\theta_i \in \left ( \xi_{b}, \xi_{b+1} \right ]$ can be done
in the following steps:
For each two adjacent bins select how many
and which letters are exchanged between the two bins and exchange the occurrences
of these letters.  Then, permute only within the
letters in a bin for all bins.  Then,
\bea
 M_{\theta, \xi} \left (\psi^n \right )
 &\stackrel{(a)}{\leq}&
 \prod_{b=1}^{A_{\xi}}
 \left \{
 \sum_{u_b=0}^{\min \left \{ \kappa_b - u_{b-1}, \kappa_{b+1} \right \}}
 \comb{\kappa_b - u_{b-1}}{u_b}
 \comb{\kappa_{b+1}}{u_b}
 \right \} \cdot
 \prod_{\beta=1}^{A_{\xi}} \kappa_{\beta}!
 \nonumber \\
 &\stackrel{(b)}{\leq}&
 \prod_{b=1}^{A_{\xi}}
 \left \{
 \sum_{u_b=0}^{\kappa_b - v_{b-1}}
 \comb{\kappa_b - v_{b-1}}{u_b} \cdot
 \sum_{v_b = 0}^{\kappa_{b+1}}
 \comb{\kappa_{b+1}}{v_b}
 \right \} \cdot
 \prod_{\beta=1}^{A_{\xi}} \kappa_{\beta}!
 \nonumber \\
 &\stackrel{(c)}{=}&
 \prod_{b=1}^{A_{\xi}}
 \left \{
 2^{\kappa_b - v_{b-1}} \cdot
 \sum_{v_b = 0}^{\kappa_{b+1}}
 \comb{\kappa_{b+1}}{v_b}
 \right \} \cdot
 \prod_{\beta=1}^{A_{\xi}} \kappa_{\beta}!
 \nonumber \\
 &\stackrel{(d)}{=}&
 2^{\kappa_1} \cdot
 \prod_{b=2}^{A_{\xi}}
 \left \{
 \sum_{v_{b-1} = 0}^{\kappa_{b}}
 \comb{\kappa_{b}}{v_{b-1}} \cdot 2^{\kappa_b - v_{b-1}}
 \right \} \cdot
 \sum_{v_{A_{\xi}} = 0}^{\kappa_{A_{\xi}+1}}
 \comb{\kappa_{A_{\xi}+1}}{v_{A_{\xi}}} \cdot
 \prod_{\beta=1}^{A_{\xi}} \kappa_{\beta}!
 \nonumber \\
 &\stackrel{(e)}{=}&
 2^{\kappa_1} \cdot
 \left \{
 \prod_{b=2}^{A_{\xi}} 3^{\kappa_b}
 \right \} \cdot
 2^{\kappa_{A_{\xi}+1}}
 \cdot
 \prod_{\beta=1}^{A_{\xi}} \kappa_{\beta}!
 ~\stackrel{(f)}{\leq}~ 3^k \cdot \prod_{\beta=1}^{A_{\xi}} \kappa_{\beta}!.
\eea
Inequality $(a)$ follows from the definition of the process above.
Some permutations within adjacent bins lead to untypical sequences, yielding an inequality.
There are up to $\min \left \{ \kappa_1, \kappa_2 \right \}$
choices of exchanging letters between bins $1$ and $2$.  (By definition
$u_0 = v_0 \dfn 0$.)
For bin $b$,
there are up to at most the number of letters in the bin not exchanged with
bin $b-1$ to exchange with bin $b+1$.
The last product represents permutations within bins after exchanges.
Inequality $(b)$ follows from $\sum a_i b_i \leq \sum a_i \cdot \sum b_i$ for
$a_i, b_i \geq 0$, and from increasing the limit of one of the sums, $(c)$ is a binomial series equality,
$(d)$ results from reorganization of terms such that the
new general term originates from the second term at index $b-1$ and the first term with
index $b$.  Binomial series relations (since $3^{\kappa_b} = (2+1)^{\kappa_b}$) lead
to $(e)$, and upper bounding the sum of all $\kappa_b$ by $k$ leads to $(f)$.
\hfill $\Box$

\section{--~~ Proof of Lemma~\ref{lemma_theorem_ub3_bounds}}
\label{ap:lemma_theorem_ub3_bounds_proof}
\renewcommand{\theequation}{D.\arabic{equation}}
\renewcommand{\theproposition}{D.\arabic{proposition}}
\renewcommand{\thelemma}{D.\arabic{lemma}}
\setcounter{equation}{0}
\setcounter{lemma}{0}
\setcounter{proposition}{0}

First, define
$\delta_i \dfn \theta_i - \rho_b \left ( \theta_i \right )$.
By definition of the $\rho_b \left (\theta_i \right )$,
$\theta_i$ and $\rho_b \left (\theta_i \right )$ must
be in the same bin.  Hence, by \eref{eq:tau_grid_spacing},
$\left | \delta_i \right | \leq 3 \sqrt{\rho_b \left ( \theta_i \right )}/
\sqrt{n}^{1+2\varepsilon}$.  Then,
\bea
 \nonumber
 \lefteqn{-n \sum_{i=k_{01}+1}^k \theta_i \log \rho_b \left ( \theta_i \right ) =
 -n \sum_{i=k_{01}+1}^k \theta_i \log \theta_i
 -n \sum_{i=k_{01}+1}^k \theta_i \log \frac{\rho_b \left ( \theta_i \right )}{\theta_i}} \\
 \nonumber
 &=&
 n H_{\theta}^{(0,1)} (X) +
 n \sum_{b = 0}^1 \varphi_b \log \varphi_b +
 n \sum_{i=k_{01}+1}^k
 \left ( \rho_b \left ( \theta_i \right ) + \delta_i \right )
 \log \left ( 1 + \frac{\delta_i}{\rho_b \left ( \theta_i \right )} \right ) \\
 \nonumber
 &\stackrel{(a)}{\leq}&
 n H_{\theta}^{(0,1)} (X) +
 n \sum_{b = 0}^1 \varphi_b \log \varphi_b +
 n \sum_{i=k_{01}+1}^k
 \left ( \rho_b \left ( \theta_i \right ) + \delta_i \right )
 \frac{\delta_i}{\rho_b \left ( \theta_i \right )} \log e \\
 &\stackrel{(b)}{\leq}&
 n H_{\theta}^{(0,1)} (X) + n
 \sum_{b = 0}^1 \varphi_b \log \varphi_b  +
 \frac{9 \log e}{n^{2\varepsilon}} \cdot
 \sum_{b \geq 2, k_b > 1} k_b
 \label{eq:ub3_term1_proof}
\eea
where $(a)$ follows from $\ln (1 + x) \leq x$, and $(b)$ is because
the total divergence from the average in any bin is $0$, the bound in
\eref{eq:tau_grid_spacing}, and since $\delta_i \neq 0$ only when $k_b > 1$.
Equation \eref{eq:ub3_term1} is proved.
Following the upper bound in \eref{eq:no_letter_i_bound} and the union bound for each
bin of $\etavec$, and
since $k_b \leq 1$ for $b > A_{\eta}$, \eref{eq:ub3_term2} is obtained.

Recall that $\ell_b \dfn \min \left \{ k_b, n \right \}$.
Then, assuming that the maximum of $\ell_b$ symbols in bin
$b$ occurred prior to any new occurrence, thus reducing the allocation to any
new symbol by $\ell_b \rho_b$,
\be
\label{eq:low_prob_cont}
 R_b \leq
 -\left ( n\varphi_b - L_b \right ) \log \rho_b -
 L_b \log \left ( \varphi_b - \ell_b \rho_b \right );~~~b = 0, 1.
\ee
While the bound is loose, it serves its
purpose well because low probability letters are unlikely to reoccur.
The minimum for \eref{eq:low_prob_cont} is attained with
$\rho_b$ in \eref{eq:rho_b_min}.  It is a valid choice of $\rho_b$ because
it leaves positive first occurrence probability after $\ell_b - 1$ first
occurrences.  Plugging \eref{eq:rho_b_min} in
\eref{eq:low_prob_cont} yields \eref{eq:low_prob_cont1}.
The following lemma is now required.
\begin{lemma}
\label{lemma_low_prob_cont}
The bound in \eref{eq:low_prob_cont1} is decreasing in $L_b$ for
$b = 0$ and also for $b=1$ if $k_1 \geq \left ( 1 + \varepsilon \right ) n^{\varepsilon}$.
\end{lemma}
As a result of Lemma~\ref{lemma_low_prob_cont}, an upper bound on $R_0$ can be derived
from \eref{eq:low_prob_cont1}
by lower bounding $L_0$ using \eref{eq:min_bin0_bound}.
Substituting \eref{eq:min_bin0_bound}, using Taylor series expansion of $\log (1 - x)$
leads to \eref{eq:bin0_cont}.

\begin{proof}{of Lemma~\ref{lemma_low_prob_cont}}
The derivative of the expression in \eref{eq:low_prob_cont1} w.r.t.\ $L_b$ is
$\log \left [ \left ( n\varphi_b - L_b \right )/ \left ( L_b \ell_b \right ) \right ]$.
It is thus negative and the function is decreasing if $n\varphi_b - L_b < L_b \ell_b$.
If $k_b \geq n$ (for either $b=0$ or $b=1$), this means that $n\varphi_b - L_b < L_b n$,
which is satisfied if $L_b > \varphi_b$.  Hence, we need to show that $L_b - \varphi_b > 0$.
Using the lower bound on $L_b$ from \eref{eq:mean_bin_bound} and the definition of $\varphi_b$,
\be
\label{eq:low_proof1}
 L_b - \varphi_b \geq k_b - \sum_{\theta_i \in \left ( \eta_b, \eta_{b+1} \right ]}
 \left [ e^{-n\theta_i} + \theta_i \right ] =
 \sum_{\theta_i \in \left ( \eta_b, \eta_{b+1} \right ]}
 \left [1 - e^{-n\theta_i} - \theta_i \right ].
\ee
The function $1 - e^{-nx} - x$ is $0$ for $x = 0$.  It increases until $x = (\ln n)/n$, and
then starts decreasing.  However, at the end of the bin $1$ region, $x = 1/n^{1-\varepsilon}$,
it still attains a positive value which goes to $1$.  Hence, since all elements
of the sum in \eref{eq:low_proof1} are positive, it must be greater than $0$.

If $n > k_b$ for $b=0$, i.e., $\theta_i \leq 1/n^{1+\varepsilon}$, we need
to prove that $\left (k_0 + 1 \right ) L_0 - n\varphi_0 > 0$.  Using the lower bound
in \eref{eq:min_bin0_bound} on $L_0$,
\be
\label{eq:low_proof2}
 \left (k_0 + 1 \right ) L_0 - n\varphi_0 \geq
 k_0 n \varphi_0 - \left ( k_0 + 1 \right )
 \comb{n}{2} \sum_{i=1}^{k_0} \theta_i^2 \geq
 \left (1-\varepsilon \right ) k_0 n \varphi_0 > 0,
\ee
where the middle inequality is since $\comb{n}{2} \sum \theta_i^2 = o \left (
n \varphi_0 \right )$.  This can be shown as follows:
Let $\theta_i \dfn \alpha_i /n^{1+\varepsilon}$ for a probability in bin $0$, where
$\alpha_i \leq 1$.  Then,
$\sum_{i=1}^{k_0} \alpha_i = \varphi_0 n^{1+\varepsilon}$.  Now,
\be
\label{eq:low_proof3}
 \comb{n}{2} \sum_{i=1}^{k_0} \theta_i^2 \leq
 \frac{1}{2n^{2\varepsilon}} \sum_{i=1}^{k_0} \alpha_i^2 \leq
 \frac{1}{2n^{2\varepsilon}} \sum_{i=1}^{k_0} \alpha_i =
 \frac{\varphi_0 n^{1-\varepsilon}}{2} = o \left ( n \varphi_0 \right ).
\ee
The second inequality is since $\alpha_i \leq 1$.

The last region is that in which $(1 + \varepsilon)n^{\varepsilon} \leq k_1 < n$.  Since
we consider bin $1$, $\theta_i \leq 1/n^{1-\varepsilon}$.  Following the same steps
as \eref{eq:low_proof1} and using the bound in \eref{eq:mean_bin_bound},
\be
 k_1 L_1 - n \varphi_1 \geq
 k_1 \cdot \left \{ \sum_{i=k_0 + 1}^{k_{01}} \left (
 1 - e^{-n\theta_i} - \frac{n\theta_i}{k_1} \right ) \right \}.
\ee
The function $1 - e^{-nx} - nx/k_1$ is $0$ for $x = 0$.  It increases until $x = (\ln k_1)/n$, and
then starts decreasing.  However, at $x = 1/n^{1-\varepsilon}$, it still
approaches at least $\varepsilon/(1+\varepsilon) > 0$ if $k_1 \geq (1+\varepsilon)n^{\varepsilon}$.
Thus, $n\varphi_1 - L_1 < k_1 L_1 = \ell_1 L_1$, and the expression in
\eref{eq:low_prob_cont1} is decreasing in $L_1$.
\end{proof}

\section{--~~ Proof of Corollary~\ref{theorem:ub3_cor1}}
\label{ap:ub3_cor1_proof}
\renewcommand{\theequation}{E.\arabic{equation}}
\renewcommand{\theproposition}{E.\arabic{proposition}}
\renewcommand{\thelemma}{E.\arabic{lemma}}
\setcounter{equation}{0}
\setcounter{lemma}{0}
\setcounter{proposition}{0}

The contributions of all $\theta_i$ such that $\theta_i \leq 1/n^{1-\varepsilon}$,
$1/n^{1+\varepsilon} < \theta_i \leq 1/n^{1-\varepsilon}$ (third and fourth terms of
\eref{eq:ub3}), and
$\theta_i \leq 1/n^{1+\varepsilon}$ (last term of \eref{eq:ub3})
considered in the two parts
of Corollary~\ref{theorem:ub3_cor1}
are bounded by the last two terms of \eref{eq:low_prob_cont1}
for $b = 01$, $b=1$, and $b=0$, respectively (recall that
\eref{eq:low_prob_cont1} also holds for $b = 01$).
Applying Lemma~\ref{lemma_low_prob_cont} to bin $b$, this expression is
decreasing in $L_b$, where for $b=1$ and $b=01$ this is provided that $k_b \geq (1+\varepsilon) n^{\varepsilon}$.
Thus a lower bound on $L_b$ yields an upper bound on these two terms.
For $k_b < (1+\varepsilon)n^{\varepsilon}$ in either case, the last two terms of
\eref{eq:ub3_c1} are
$O \left ( n^{2\varepsilon} \log n \right )$ because $\varphi_b \leq k_b/n^{1-\varepsilon}
< (1 + \varepsilon)/n^{1-2\varepsilon}$.

Now, $L_b$ is lower bounded similarly for $b=0,1,01$.
Let $\theta_M$ denote the maximal probability
in bin $b$, and denote the probability of letter $i$ in bin $b$
by $\theta_i = \alpha_i \theta_M$, $\alpha_i \leq 1$.  Using \eref{eq:mean_bin_bound},
\bea
 \nonumber
 L_b &\geq& k_b - \sum_i e^{-n\theta_i} =
 \sum_i \left ( 1 - e^{-n\theta_M \alpha_i} \right ) \\
 &\stackrel{(a)}{\geq}&
 \sum_i \alpha_i \left (1 - e^{-n\theta_M} \right )
 \stackrel{(b)}{=} \frac{\varphi_b}{\theta_M}
 \left ( 1 - e^{-n\theta_M} \right )
 \label{eq:ub3_b0b1_letters}
\eea
where $(a)$ follows from $1 - x^{\alpha} \geq \alpha (1 - x)$ for
$0 < \alpha \leq 1$ and $0 < x \leq 1$, because
$1 - x^{\alpha} - \alpha +\alpha x$ equals $1-\alpha$ for
$x = 0$, $0$ for $x=1$, and is decreasing between $x=0$ and $x=1$.
Equality $(b)$ follows from $\varphi_b = \sum \theta_i = \sum \alpha_i \theta_M
\Rightarrow \sum \alpha_i = \varphi_b / \theta_M$.
Using Taylor series expansion,
\be
 \label{eq:bin1_L1_bound}
 \left ( n \varphi_b - L_b \right ) \log \ell_b +
 n \varphi_b \cdot h_2 \left ( \frac{L_b}{n\varphi_b} \right ) \leq
 n\varphi_b \log \left (  \ell_b \right ) +
 L_b \log \frac{ n \varphi_b e}{\ell_b L_b}.
\ee
Substituting \eref{eq:ub3_b0b1_letters} to lower bound $L_b$ for $b = 1$ and $b = 01$
($\theta_M = 1/n^{1-\varepsilon}$),
\bea
 \nonumber
 \lefteqn{\left ( n \varphi_b - L_b \right ) \log \ell_b +
 n \varphi_b \cdot h_2 \left ( \frac{L_b}{n\varphi_b} \right )}\\
 \label{eq:bin1_worst_case}
 &\leq&
 n \varphi_b \log \ell_b +
 \varphi_b n^{1-\varepsilon} \log \frac{e n^{\varepsilon}}{\ell_b} +
 \Theta \left ( \varphi_b n^{1-\varepsilon} e^{-n^{\varepsilon}}\right ) =
 O \left ( n \varphi_b \log n \right ),
\eea
concluding the proof for Part I.

For $b=0$, the second term of
\eref{eq:bin0_cont} is increasing in
$\sum \theta_i^2$.  Thus, using \eref{eq:low_proof3},
\be
\label{eq:bin0_wc1}
 \left ( \frac{n^2}{2} \sum_{i=1}^{k_0} \theta_i^2 \right )
 \log \frac{2 e \cdot \varphi_0 \cdot \min \left \{k_0, n \right \}}
 {n \sum_{i=1}^{k_0} \theta_i^2} \leq
 \frac{\varphi_0 n^{1-\varepsilon}}{2} \log \left ( 2 e n^{1+\varepsilon} \right ).
\ee
For the other statement of Part II, first, if $\forall \theta_i \leq 1/n^{1+\varepsilon}$,
also $\theta_i \leq 1/n^{\mu + \varepsilon}$, then,
\be
 \left ( \frac{n^2}{2} \sum_{i=1}^{k_0} \theta_i^2 \right )
 \log \frac{2 e \varphi_0 \ell_0}{n \sum_{i=1}^{k_0} \theta_i^2} \leq
 \frac{\varphi_0}{2} n^{2-\mu-\varepsilon} \log \left ( 2 e n^{\mu+\varepsilon} \right )
 = O \left ( n^{2-\mu-\varepsilon} \log n \right )
\ee
by the same arguments of \eref{eq:bin0_wc1}.  Otherwise,
\be
 \label{eq:ub3_cor1_B}
 \frac{n^2}{2} \sum_{\theta_i \leq 1/n^{\mu + \varepsilon}} \theta_i^2 \leq
 \frac{\varphi_{0\mu}}{2} n^{2-\mu-\varepsilon},
\ee
where $\varphi_{0\mu} \dfn \sum_{\theta_i \leq 1/n^{\mu + \varepsilon}} \theta_i$, and also
$\sum_{i=1}^{k_0} \theta_i^2 > 1/n^{2\mu + 2\varepsilon}$, because $\exists \theta_i > 1/n^{\mu+\varepsilon}$
in bin $0$.  Therefore,
\be
 \left ( \frac{n^2}{2} \sum_{\theta_i \leq 1/n^{\mu+\varepsilon}}
 \theta_i^2 \right )
 \log \frac{2 e \varphi_0 \ell_0}{n \sum_{j=1}^{k_0} \theta_j^2} \leq
 \frac{\varphi_{0\mu}}{2} n^{2-\mu-\varepsilon} \log \left ( 2 e n^{2\mu+2\varepsilon} \right )
 = O \left ( n^{2-\mu-\varepsilon} \log n \right ),
\ee
where $\ell_0 \leq n$ and $\varphi_0 \leq 1$ are used.
\hfill $\Box$

\section{--~~ Proof of Lemma~\ref{lemma_low_prob_boundary}}
\label{ap:lemma_low_prob_boundary_proof}
\renewcommand{\theequation}{F.\arabic{equation}}
\renewcommand{\theproposition}{F.\arabic{proposition}}
\renewcommand{\thelemma}{F.\arabic{lemma}}
\setcounter{equation}{0}
\setcounter{lemma}{0}
\setcounter{proposition}{0}

Four regions of $\theta_i$ are considered:  $\theta_i \leq \mu_j/n^{1-\varepsilon}$,
$j =1,2$, and $\theta_i > \mu_j/n^{1-\varepsilon}$, $j = 3,4$,
where $\{\mu_j\} = \left ( \vartheta^-, 1, 1, \vartheta^+ \right )$, respectively.  Let
$\{\nu_j\} = \left (\gamma^-, \gamma^+, \gamma^-, \gamma^+ \right )$, respectively,
where $\vartheta^- < \gamma^- < 1$ and $1 < \gamma^+ < \vartheta^+$.
Now, let
\be
 \label{eq:event_A_new}
 \eventA = \left \{ x^n : \exists \hat{\theta}_i;
 \hat{\theta}_i > \frac{\nu_j}{n^{1-\varepsilon}}~\mbox{for}~
 \theta_i \leq \frac{\mu_j}{n^{1-\varepsilon}};
 j=1,2,~\mbox{or}~
 \hat{\theta}_i \leq \frac{\nu_j}{n^{1-\varepsilon}}, ~\mbox{for}~
 \theta_i > \frac{\mu_j}{n^{1-\varepsilon}};
 j=3,4
 \right \}
\ee
be the event that for $\theta_i$ in one of the four regions defined above
there exists an empirical ML estimate on the other side of the probability interval,
that is separated from the boundary of the region of $\theta_i$ by at least a complete interval between points in
$\left ( \vartheta^-, \gamma^-, 1, \gamma^+, \vartheta^+ \right )/n^{1-\varepsilon}$.
By typicality arguments and the union bound
\be
 \label{eq:event_A_new_prob}
 P_{\theta} \left (\eventA \right ) \leq
 n \cdot k_{\theta_i > n^{-3}} \cdot
 2^{-n \min_{\eventA} D \left (\hat{\theta}_i ||\theta_i \right )} + \frac{1}{2 \gamma^- n^{1+\varepsilon}},
\ee
where
the additional term bounds the probability of re-occurrence $\gamma^- n^{\varepsilon}$ or more times
of any letter with
$\theta_i \leq 1/n^3$ using \eref{eq:mean_recur_bound0}, the bound
in \eref{eq:ub3_cor1_B} with $\mu + \varepsilon = 3$, and
Markov's inequality.
Then, the union bound on the number of remaining letters
(where $k_{\theta_i > n^{-3}}$ denotes the total letters with
$\theta_i > 1/n^3$) and the number of types (at most $n$)
produces the first term.  If $\eventA$ occurs in region $j$,
\be
 \label{eq:min_divergence_correct_term}
 D \left (\hat{\theta}_i ||\theta_i \right ) \geq
 \frac{\nu_j}{n^{1-\varepsilon}} \log \frac{\nu_j}{\mu_j} +
 \left ( 1 - \frac{\nu_j}{n^{1-\varepsilon}} \right )
 \log \frac{n^{1-\varepsilon} - \nu_j}{n^{1-\varepsilon} - \mu_j} \geq
 \frac{1}{n^{1-\varepsilon}} \left [
 \nu_j \log \frac{\nu_j}{\mu_j} +
 \left ( \mu_j - \nu_j \right ) \log e
 \right ]
\ee
where the second inequality follows Taylor expansion.
The values of $\gamma^-$ and $\gamma^+$ can be optimized to maximize
the divergence in \eref{eq:min_divergence_correct_term} by trading off between $j=1$ and $j=3$
for $\gamma^-$ and between $j=2$ and $j=4$ for $\gamma^+$.  This yields
\be
 \gamma^{\pm} = \frac{\vartheta^{\pm} - 1}{\ln \vartheta^{\pm}}
\ee
where $\pm$ is used to denote both cases.  Plugging these choices of $\gamma^{\pm}$, if $\eventA$
occurs
\be
 D \left (\hat{\theta}_i ||\theta_i \right ) \geq
 \frac{1}{n^{1-\varepsilon}} \left [
 \min \left \{
 \frac{\vartheta^{\pm} - 1}{\ln \vartheta^{\pm}}
 \log \frac{\vartheta^{\pm} - 1}{e \cdot \ln \vartheta^{\pm}} + \log e
 \right \}
 \right ]
\ee
where the minimum is taken between the value of the expression for $\vartheta^-$ and for $\vartheta^+$.
Hence,
\bea
 \label{eq:event_A_new_prob1}
 P_{\theta} \left (\eventA \right )
 &\leq& \varepsilon'_n \dfn
 n \cdot k_{\theta_i > n^{-3}} \cdot
 e^{-f \left ( \vartheta^-, \vartheta^+ \right ) n^{\varepsilon}} +
 \frac{\ln \vartheta^-}{2 (\vartheta^- - 1) n^{1+\varepsilon}},~~\mbox{where} \\
 f \left ( \vartheta^-, \vartheta^+ \right ) &\dfn&
 \min \left \{
 \frac{\vartheta^{\pm} - 1}{\ln \vartheta^{\pm}}
 \ln \frac{\vartheta^{\pm} - 1}{e \cdot \ln \vartheta^{\pm}} + 1
 \right \}.
\eea
Specifically, choices of $\vartheta^- = e^{-5.5} \approx 0.004$
and $\vartheta^+ = e^{1.4} \approx 4.06$ result in
$\gamma^- \approx 0.18$, $\gamma^+ \approx 2.18$,
$f \left ( \vartheta^-, \vartheta^+ \right ) > 0.5$, and an upper bound
of $2.77/n^{1+\varepsilon}$ on the
last term of \eref{eq:event_A_new_prob1}.

Let $F$ denote the Bernoulli event of whether event $\eventA$ occurs.  Then,
\bea
 \nonumber
 H_{\theta} \left ( Z^n~|~\Psi^n \right ) &\leq&
 H_{\theta} \left ( Z^n, F ~|~ \Psi^n \right )
 ~=~
 H_{\theta} \left ( Z^n ~|~ \Psi^n, F \right ) + H_{\theta} \left ( F ~|~ \Psi^n \right ) \\
 \nonumber
 &\leq&
 P_{\theta} \left ( \bar{\eventA} \right )
 H_{\theta} \left ( Z^n ~|~ \Psi^n, \bar{\eventA} \right ) +
 P_{\theta} \left ( \eventA \right )
 H_{\theta} \left ( Z^n ~|~ \Psi^n, \eventA \right ) +
 H_{\theta} \left (F \right )\\
 \label{eq:correction_term_lb_proof}
 &\stackrel{(a)}{\leq}&
 \log \comb{k_{\vartheta}^- + k_{\vartheta}^+}{k_{\vartheta}^+} +
 \varepsilon'_n n +
 o(\varepsilon'_n n),
\eea
where $(a)$ follows since
given $\bar{\eventA}$, the only uncertainty about $Z^n$ is for indices
for which $\hat{\theta}_i \in \left (\gamma^-/n^{1-\varepsilon}, \gamma^+/n^{1-\varepsilon} \right ]$, because in all
other regions it is guaranteed that $\hat{\theta}_i$ is on the correct
side of $1/n^{1-\varepsilon}$, thus there is no uncertainty about the value of $z_{\ell}$
corresponding to such $\psi_{\ell}$.  The only symbols for which it is possible to have
$\hat{\theta}_i \in \left (\gamma^-/n^{1-\varepsilon}, \gamma^+/n^{1-\varepsilon} \right ]$
are the $k_{\vartheta}^- + k_{\vartheta}^+$ letters with
$\theta_i \in \left (\vartheta^-/n^{1-\varepsilon}, \vartheta^+/n^{1-\varepsilon} \right ]$.  The uncertainty
in $Z^n$ is choosing which such symbols correspond to $z = 1$, and the
worst case is when the total
possible choices of $k_{\vartheta}^+$ out of $k_{\vartheta}^- + k_{\vartheta}^+$ are uniformly distributed.
The second term is since $H_{\theta} \left ( Z^n ~|~ \Psi^n, \eventA \right ) \leq n$ for the
Bernoulli process $Z^n$.
\hfill $\Box$

\section*{Acknowledgments}

The author gratefully acknowledges associate editor, Wojciech
Szpankowski, and an anonymous reviewer for comments that led to significant
improvements in the exposition in this paper.


\begin{thebibliography}{999}

\bibitem{aberg97}J. {\AA}berg, Y. M. Shtarkov, and B. J. M. Smeets, ``Multialphabet
coding with separate alphabet description,'' in \emph{Proc.\ of Compression and
Complexity of Sequences\/}, pp. 56-65, Jun.\ 1997.

\bibitem{cover91}  T. M. Cover and J. A. Thomas, \emph{Elements of
Information Theory\/}, John Wiley \& Sons, 1991.

\bibitem{cizar81}  I. Csiszar and J. Korner, \emph{Information Theory: Coding
Theorems for Discrete Memoryless Systems.\/}, Academic Press, New
York, 1981.

\bibitem{davisson73} L. D. Davisson,
``Universal noiseless coding,'' \emph{IEEE Trans. Inform.
Theory\/}, vol. IT-19, no. 6, pp. 783-795, Nov.\ 1973.

\bibitem{gemelos06} G. M. Gemelos and T. Weissman,
``On the entropy rate of pattern processes,'' \emph{IEEE Trans.
Inform. Theory\/}, vol.\ 52, no.\ 9, pp.\ 3994-4007, Sep.\ 2006.

\bibitem{jevtic02} N. Jevti\'c, A. Orlitsky, N. Santhanam,
``Universal compression of unknown alphabets,'' in
\emph{Proc.\ of 2002 IEEE International Symposium on
Information Theory\/}, Lausanne, Switzerland, p. 320, Jun.\ 30-Jul.\
5, 2002.


\bibitem{orlitsky04} A. Orlitsky, N. P. Santhanam, and J. Zhang,
``Universal compression of memoryless sources over unknown
alphabts,'' \emph{IEEE Trans. Inform. Theory\/}, vol. 50, no. 7,
pp. 1469-1481, Jul.\ 2004.



\bibitem{orlitsky06} A. Orlitsky, N. P. Santhanam, K. Viswanathan, and J. Zhang,
``Limit results on pattern entropy,''
\emph{IEEE Trans. Inform. Theory\/}, vol.\
52, no.\ 7, pp.\ 2954-2964, Jul.\ 2006.

\bibitem{rissanen84}  J. Rissanen, ``Universal coding, information, prediction,
and estimation,'' \emph{IEEE Trans. Inform. Theory\/}, vol. IT-30,
no. 4, pp. 629-636, Jul.\ 1984.

\bibitem{shamir05} G. I. Shamir,
``Applications of coding theory to universal lossless source
coding performance bounds,'' in \emph{DIMACS Series in Discrete Mathematics
and Theoretical Computer Science\/}, A. Ashikhmin, A. Barg, Eds. American
Mathematical Society, vol. 68, pp. 21-55, 2005.

\bibitem{shamir04} G. I. Shamir,
``On the MDL principle for i.i.d.\ sources with large alphabets,''
\emph{IEEE Trans. Inform. Theory\/}, vol.\ 52, no.\ 5, pp. 1939-1955, May 2006.

\bibitem{shamir03} G. I. Shamir,
``Universal lossless compression with unknown alphabets - the
average case'', \emph{IEEE Trans. Inform. Theory\/},
vol.\ 52, no.\ 11, pp.\ 4915-4944, Nov.\ 2006.

\bibitem{shamir07} G. I. Shamir,
``Patterns of i.i.d. sequences and their entropy - Part II: bounds for some
distributions,'' sumbitted to \emph{IEEE Trans. Inform. Theory\/}.

\bibitem{shamir03a} G. I. Shamir and L. Song,
``On the entropy of patterns of i.i.d. sequences,''
in \emph{Proc.\ of The 41st Annual Allerton Conference
on Communication, Control, and Computing\/}, Monticello, IL, U.S.A.,
pp. 160-169, Oct.\ 1-3, 2003.

\bibitem{shamir04c} G. I. Shamir,
``A new redundancy bound for universal lossless compression of
unknown alphabets,'' in \emph{Proc.\ of The 38th Annual
Conference on Information Sciences and Systems\/}, Princeton,
New-Jersey, U.S.A., pp.\ 1175-1179, Mar.\ 17-19, 2004.

\bibitem{shamir04d} G. I. Shamir,
``Sequential universal lossless techniques for compression of
patterns and their description length,'' in \emph{Proceedings of
The Data Compression Conference\/}, Snowbird, Utah, U.S.A., pp.\
419 - 428, Mar.\ 23-25, 2004.

\bibitem{shamir04a} G. I. Shamir,
``Sequence-patterns entropy and infinite alphabets,'' in
\emph{Proc.\ of The 42nd Annual Allerton Conference on
Communication, Control, and Computing\/}, Monticello, IL, U.S.A.,
pp.\ 1458-1467, Sep.\ 29 - Oct.\ 1, 2004.

\bibitem{shamir05itw} G. I. Shamir, ``Bounds on the entropy of patterns of i.i.d.\
sequences,'' in \emph{Proc.\ of the IEEE Information Theory Workshop on Coding and
Complexity\/}, Rotorua, New Zealand, pp.\ 202-206, Aug.\ 29-Sep.\ 1, 2005.

\bibitem{shamir06} G.\ I.\ Shamir, ``On some distributions and their pattern entropies,''
in \emph{Proc.\ of the 2006 IEEE International Symposium on
Information Theory\/}, Seattle, Washington, U.S.A., pp.\ 2541-2545, Jul.\ 9-14,
2006.

\bibitem{shtarkov95}  Y. M. Shtarkov, T. J. Tjalkens and F. M. J. Willems,
``Multi-alphabet universal coding of memoryless sources,'' \emph{Problems of
Information Transmission\/}, vol. 31, no. 2, pp 20-35, Apr.\-Jun., 1995.


\end{thebibliography}
\end{document}